\documentclass[aps,prb,showpacs,twocolumn,longbibliography, superscriptaddress]{revtex4-1}

\usepackage[dvips]{graphicx}
\usepackage{amsmath,amssymb}
\usepackage{amsfonts}
\usepackage[table,usenames,dvipsnames]{xcolor}
\usepackage{bbold}
\usepackage[normalem]{ulem}

\begin{document}
\newcommand{\Eq}[1]{Eq.~\eqref{#1}}
\newcommand{\Eqs}[1]{Eqs.~\eqref{#1}}

\newcommand{\phiL}{\phi_L}
\newcommand{\phiR}{\phi_R}
\newcommand{\phiBB}{\phi^{B}}
\newcommand{\phiCC}{\phi^{C}}

\newcommand{\IJab}{\hat I_c}
\newcommand{\CJab}{\hat C_J}
\newcommand{\IJbb}{I_c^{B}}
\newcommand{\CJbb}{C_J^{B}}

\newcommand{\Efl}{E_{\text{fl}}}
\newcommand{\sech}{\;\text{sech}} 

\newcommand{\uGHz}{\,\text{GHz}}
\newcommand{\uzJ}{\,\text{zJ}}
\newcommand{\umA}{\,\mu\text{A}}
\newcommand{\upH}{\,\text{pH}}
\newcommand{\ufF}{\,\text{fF}}

\title{Asynchronous Reversible Computing Unveiled Using Ballistic Shift Registers}

\author{K.D.~Osborn}
\email{corresponding author, osborn@lps.umd.edu (alternatively kosborn@umd.edu)}
\affiliation{The Laboratory for Physical Sciences at the University of Maryland, 
College Park, MD 20740, USA}
\affiliation{The Joint Quantum Institute, University of Maryland, 
College Park, MD 20742, USA}

\author{W.~Wustmann}
\affiliation{The Laboratory for Physical Sciences at the University of Maryland, 
College Park, MD 20740, USA}

\begin{abstract}
Reversible logic can provide lower switching energy costs relative to all 
irreversible logic, including those developed by industry in semiconductor 
circuits, however, more research is needed to understand what is possible. 
Superconducting logic, an exemplary platform for both irreversible and 
reversible logic, uses flux quanta to represent bits, and the reversible 
implementation may switch state with low energy dissipation relative to the 
energy of a flux quantum. Here we simulate reversible shift register gates 
that are ballistic: their operation is powered by the input bits alone. 
A storage loop is added relative to previous gates as a key innovation, 
which bestows an asynchronous property to the gate such that input bits 
can arrive at different times as long as their order is clearly preserved. 
The shift register represents bit states by flux polarity, both in the stored 
bit as well as the ballistic input and output bits. Its operation consists of 
the elastic swapping of flux between the stored and the moving bit. 
This is related to a famous irreversible shift register, developed prior to 
the advent of superconducting flux quanta logic (which used irreversible gates). 
In the base design of our ballistic shift register (BSR) there is one 
1-input and 1-output port, but we find that we can make other asynchronous 
ballistic gates by extension. For example, we show that two BSRs operate in 
sequence without power added, giving a 2-bit sequential memory. 
We also show that a BSR with two input and two output ports allows a bit state 
to be set from one input port and then conveyed out on either output port. 
The gate constitutes the first asynchronous reversible 2-input gate. 
Finally, for a better insight into the dynamics, we introduce a collective 
coordinate model. We find that the gate can be described as motion in two 
coordinates subject to a potential determined by the input bit and initial 
stored flux quantum. Aside from the favorable asynchronous feature, 
the gate is considered practical in the context of energy efficiency, 
parameter margins, logical depth, and speed.
\end{abstract}

\maketitle

\section{Introduction}

Modern superconducting digital logic \cite{Holmes2013, SolETAL2017} 
uses Single Flux Quanta (SFQ) as bits, 
and has been developed through research programs since the 80s \cite{LikMukSem1985}. 
Its circuits are built from inductors and resistively-shunted Josephson junctions. 
Today, superconducting logic is primarily demonstrated in three types: 
RSFQ descendants \cite{LikMukSem1985, LikSem1991, KirETAL2011, VolETAL2013, TanETAL2015},  
AQFP \cite{Harada1987_QFP,  YoshikawaETAL2021}, and RQL \cite{HerrETAL2011}.  
The logic gates use conditional activation of SFQ over potential energy barriers 
with an energy cost $\gg k_B T$.
The energy cost is bound by a minimum, $\ln(2) k_B T$ per bit, 
due to irreversible loss of information in logically irreversible gate operations \cite{Landauer1961}. 

Arguably, the earliest influential SFQ work \cite{LikSem1991_App2}
is a proposal for a ``Flux Shuttle'' by the Bell Labs team of Anderson, Dynes, and Fulton\cite{FulDynAnd1973_fluxshuttle, FulDun1973_expfluxshuttle}. 
The device shows that an SFQ can be advanced within the circuit from one storage location to the next by an external current pulse; since the SFQ can represent a bit state, the shuttle demonstrates a precursor to a shift register. 
This is similar to a shift register in RSFQ \cite{MukhanovETAL1991, Mukhanov1993}, 
but in RSFQ the shift forward is caused by the arrival of another SFQ (a clock SFQ) rather than an external current pulse. 
Both structures are not thermodynamically reversible due to the activated dynamics, 
which relies on resistive elements for damping. 
This damping is needed in irreversible logic to allow the circuit to quickly 
reach a new steady state. 

Here we introduce and study ballistic shift registers (BSRs). 
Due to their reversible design (time-reversal symmetry), 
they can in principle incur an energy cost $< \ln(2) k_B T$ per bit operation. 
The devices operate without shunt resistors, and a key feature is that the inertia of the input bits 
solely powers an operation that leaves the device near a new steady state after the operation. 
Unlike the `asymmetric' bit-state encoding in RSFQ, based on SFQ presence versus absence, 
BSRs use the two degenerate flux states (polarities) to represent the bit states, 
both for the moving bit (at input and output) and the stored bit. 
Since the data SFQ travels freely and the gate is unpowered, 
it is considered ``ballistic''. 
In the gate, the moving SFQ interacts with an SFQ stored by an inductor. 
Nonlinear dynamics allow an input bit to transform the stored state to the input bit state and also to carry away the input bit energy as an output SFQ with the previously stored bit state. 
This constitutes a shift register operation from the forward-scattering dynamics of an SFQ. 
To accomplish the near-ideal reversible dynamics, the gates utilize
capacitive shunts in the JJs within the gate circuit.

The BSR gates expand upon previous work \cite{WusOsb2020_RFL, WusOsb2020_CNOT, OsbWus2018_CNOT}
on a reversible logic named Reversible Fluxon Logic (RFL),
which uses ballistic SFQs moving along long Josephson junctions (LJJs),
where the SFQs are spatially extended and are referred to as fluxons. 
Ballistic RFL gates always use LJJs in pairs named bit lines, 
that carry ballistic bits into and out of the gate. 
In that work the ballistic gates for a CNOT \cite{WusOsb2020_CNOT} 
and NSWAP \cite{WusOsb2020_RFL} require synchronized input bits. 
RFL is distinct from the reversible circuits 
of parametric quantron \cite{parametricquantron1985},
nSQUID \cite{SemDanAve2003, RenSemETAL2009, RenSem2011} 
and AQFP \cite{RQFPgate}, 
because those circuits adiabatically apply power 
from a clock to execute the gate operation. 
In contrast, in ballistic RFL gates the bits use no clock reference. 
The RFL CNOT is not fully ballistic and uses clock SFQ. 

BSR gates, as a new development of RFL, are asynchronous. 
Asynchronous ballistic gates require a stored state such that the ballistic bit can interact with it \cite{Frank2017}.
By definition, asynchronous reversible logic 
allows an arbitrary delay time between input bits, as long as it exceeds a minimum delay time. 
Asynchronous logic thus avoids the requirement of synchronized bits. 
Moreover, it has the potential for the sequential operation of multiple gates without clocking, and this provides an architectural advantage over clocked gates, such as typical gates in RSFQ. 
In this work, we show simulated operations of the first asynchronous reversible gates, 
a sequential BSR and a BSR with multiple input ports. 
The latter gate allows one to write a bit state on one bit line and then transfer it to a second bit line, which is a (different) 2-bit shift operation.  

As the BSR operation is unpowered, it relies on the free scattering dynamics of the input fluxon. The dynamics depend on the difference between the input and stored bit states, such that there are 2 dynamical types. We chose a combination of dynamical types that are favorable in terms of parameter margins.
If the input and stored bit states differ, the dynamics are similar to a 1-bit NOT gate, 
which uses a resonance for polarity inversion \cite{WusOsb2020_RFL}. 
If the input and stored bit states are equal, the dynamics are of transmission type, 
which are simpler than a previously studied ID resonance gate \cite{WusOsb2020_RFL}. 
This combination improves the parameter margins compared with the 2-bit synchronous IDSN gate \cite{WusOsb2020_CNOT}, which combines the NOT and ID dynamical types. 

This article is organized as follows: 
In Sec.~\ref{sec:SR} we present a 1-input BSR, analyze its steady states,
and discuss the operations of the BSR for both a 1-bit 
and multi-bit serial shift register. 
A 2-input BSR is introduced in Sec.~\ref{sec:2-inputSR}, and 
Sec.~\ref{sec:margins} gives an overview of the operation margins.
In Sec.~\ref{sec:timing} we investigate the 
fluxon delay times, and give estimates for 
the timing uncertainty (jitter), 
induced either by the gate or by thermal fluctuations.
In Sec.~\ref{sec:CCM}, 
we analyze the BSR dynamics by means of a
collective coordinate model (CCM), 
which is shown to quantitatively describe the BSR dynamics and 
to help with the interpretation of the gate dynamics. 
A discussion section, Sec.~\ref{sec:discussion}, 
includes technical findings on: 
energy efficiency, speed, energy-delay product, and logical depth.

\section{BSR circuit and operation}\label{sec:SR}

The Anderson-Dynes-Fulton (ADF) flux shuttle 
is an important historical step in the development of SFQ digital electronics, 
because it interprets SFQ as bits and it is closely related to RSFQ shift registers 
even though it predates RSFQ by approximately a decade \cite{LikSem1991_App2}. 
In the ADF flux shuttle, the SFQ can be localized in a potential well 
generated by device geometry or magnetic fields, and can be 
moved forward to the next potential well by current pulses \cite{FulDun1973_expfluxshuttle}.

In contrast, RSFQ shift registers \cite{MukhanovETAL1991, Mukhanov1993}
use DC current bias and SFQ clock signals to forward the bits.
As is standard in RSFQ, a data SFQ represents the logic 1-state, while 
the 0-state is represented by an absence of flux at the same position.
In these shift registers,
a clock SFQ arriving near a data storage cell
causes a JJ in the storage cell to switch phase by $2\pi$ if that cell contains 
a data SFQ. 
In this case, the data SFQ will be shifted to an adjacent cell. 
The clock SFQ will progress past the storage cell,
regardless of the presence of a data SFQ in it. 

In the RSFQ shift registers, and RSFQ gates generally \cite{LikSem1991}, 
the power to move SFQ comes from current biases.
As the JJs in RSFQ circuits are critically damped and biased 
with a bias current $I_{\text{bias}} \lesssim I_c$
near their critical current $I_c$, the $2\pi$-phase switching
is accompanied by an energy dissipation of $\lesssim I_c \Phi_0$,
where $\Phi_0$ is the flux quantum. 
This dissipated energy is of the same order of magnitude 
as the energy of the logic 1-state. 
For context, note that a fluxon bit in an LJJ can be related to the SFQ energy 
in a typical digital cell: 
the typical bit energy for an SFQ is $\sim J_c d^2 \Phi_0$ with JJ critical 
current density $J_c$ and JJ diameter $d$. 
The energy of a fluxon in a long Josephson 
junction is $\sim J_c w \lambda_J \Phi_0$ for a long junction 
with width $w$ (the `short' dimension) and Josephson penetration depth $\lambda_J$,
where the latter determines the fluxon length.

Reversible fluxon logic (RFL) represents the bit states 0 and 1 
by the two possible polarities $\sigma=\pm 1$ of a fluxon, 
corresponding to the sign of its flux $\pm \Phi_0$
and denoted as fluxon ($+$) and antifluxon ($-$). 
Switching between the degenerate bit states (polarity inversion) and other 
logic operations may be achieved in ballistic gates 
\cite{WusOsb2020_RFL, WusOsb2020_CNOT, Liuqi2019}, 
which are undriven and solely powered by the energy of the input fluxon(s).
Previous ballistic reversible gates of RFL
had been designed without internal state memory, 
implying that the operation of a multi-bit gate requires synchronous input bits.
Although specialized store-and-launch gates \cite{OsbWus2018_CNOT, WusOsb2020_CNOT}
have been designed for the purpose of synchronization (and routing), 
others have advocated for `asynchronous' ballistic reversible gates \cite{Frank2017}.

Asynchronous multi-bit gates have the advantage that the timing of the input bits 
no longer needs to be precise. 
They merely have to arrive with a minimum delay time between them 
to allow quiescence before the next arrival. 
Asynchronous ballistic gates require an internal state, 
by which an interaction between subsequent input bits is mediated. 
In order to generate output that depends on the input of a previous scattering,
the internal state has to: be changeable by an input bit,
and determine the output state of the ballistic scattering.

The storage cell of the ballistic shift register (BSR) provides exactly such 
functionality, in that it can store a bit state 
in form of a flux quantum with positive or negative flux orientation $S=\pm 1$.
As we describe below, 
the stored state $S$ may change during the scattering dynamics, 
depending on the input fluxon state $\sigma$. 
The scattering type (output fluxon state) 
in turn depends on the current value of $S$, 
but is independent of the input timing, regardless of what input ports are used.

\subsection{Circuit}

\begin{figure}[b]
\includegraphics[width=8.8cm]{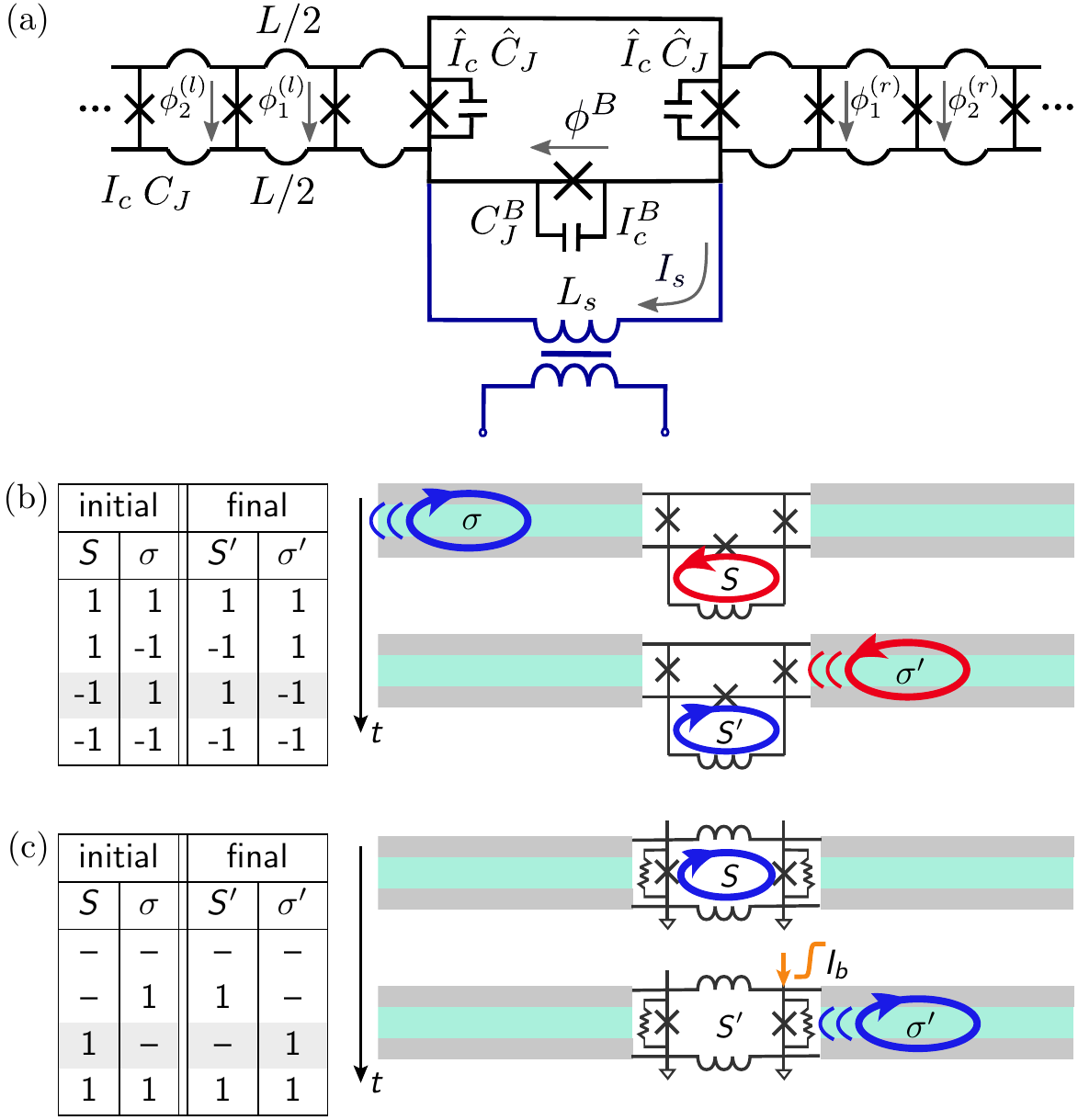}
\caption{
(a) Circuit schematic for a 1-input BSR,
consisting of one input and one output LJJ (3 cells shown for each) 
connected by a circuit interface, which is made from
three capacitance-shunted JJs,
where two of these are left and right `termination JJs' with $(\CJab,\IJab)$
and the third is the `rail JJ' with $(\CJbb, \IJbb)$. 
An inductor $L_s$ in parallel with the rail JJ is made to store one SFQ, 
where a clockwise (counterclockwise) circulating current $I_s$ corresponds 
to flux state $S=1$ ($S=-1$) and rail phase $\phiBB \approx 2\pi S$.
The parameters of the interface and storage cell are set to enable energy-efficient 
forward-scattering from one LJJ to the other 
for all combinations of stored flux state $S=\pm 1$
and input flux state $\sigma=\pm 1$, for a range of input velocities.
External circuitry coupled inductively to $L_s$
may be optionally used to assist initialization of the BSR (loading), giving $|S|=1$.
(b) Operation table for 1-input BSR characterized by 
SWAP-operation between stored and moving flux states: ($S',\sigma'$)=SWAP($S,\sigma$).
Schematical illustration of BSR operation is shown for third row of table, 
with $S=-1$, $\sigma=1$, and resulting $S'=1$, $\sigma'=-1$. 
The LJJ trilayer is illustrated using grey for superconductor 
and blue for tunneling barrier. 
(c) Operation table for the pioneering ADF flux shuttle \cite{FulDynAnd1973_fluxshuttle},
which is based on the presence of an SFQ as the bit state. 
Schematic illustration of the shuttle operation is shown for third row of table, 
where a fluxon settles as a static SFQ in a storage cell due to damping 
(shown as shunt resistors), and is subsequently released as a fluxon 
by application of a bias current $I_b$.
}
\label{fig:circuit_SR}
\end{figure}

The most basic BSR is the 2-port circuit shown in Fig.~\ref{fig:circuit_SR}(a).
It consists of a storage cell,
one LJJ each for input and output, and a special interface cell between them. 
The LJJs form a part of the gate and also  
serve as fluxon in- and output channels.
The side arms of the interface cell are formed by two JJs with parameters
$(\CJab,\IJab)$. Each of them terminates one of the LJJs and thus they are 
referred to as `termination JJs'.
The upper rails of the LJJs are joined by a negligible inductance
(on the upper side of the interface cell), 
while the lower rails are connected by the so-called `rail JJ' 
(of the interface cell), with parameters $(\CJbb,\IJbb)$.
The rail JJ is also part of the 
storage cell which is closed by a parallel inductor $L_s$.
Given suitable parameters ($2\pi L_s \IJbb/\Phi_0 \gtrsim 5$), 
the storage cell can store 
one bit of information in form of a steady circulating current.
A clockwise (counterclockwise) circulating current corresponds 
to a positive (negative) flux orientation $S=1$ ($S=-1$),
and rail-JJ phase $\phiBB \approx 2\pi S$.

Similar to the 1-bit ballistic RFL gates, the NOT or ID,
the interface of the BSR is designed to enable 
forward-scattering, i.e., 
starting from a fluxon entering 
on one LJJ and resulting in a fluxon exiting on the other LJJ. 
By making the in- and output LJJs sufficiently long ($\gtrsim 10 \lambda_J$) 
we ensure that the ballistic fluxons can move freely.
The ballistic gates require specific parameter values in the interface to achieve the operation. 
The ballistic scattering involves the temporary breaking of the fluxon at the gate interface and a short oscillation of an interface mode. 
In previous 1-bit ballistic RFL gates, i.e., the NOT and ID, 
the polarity of the exiting fluxon is determined by the polarity of the incoming
fluxon alone. 
In contrast, the output bit of the BSR is dependent on the stored bit state $S$.
The ballistic scattering dynamics generates the regular operations 
of a shift register, summarized in the table of Fig.~\ref{fig:circuit_SR}(b).
One of the four possible operations of the BSR is sketched in 
Fig.~\ref{fig:circuit_SR}(b).
In comparison with the operation of the ADF flux shuttle 
\cite{FulDynAnd1973_fluxshuttle, FulDun1973_expfluxshuttle},
Fig.~\ref{fig:circuit_SR}(c),
it uses no external drive power to advance the stored SFQ from the storage cell. 
Instead, the incoming bit state is swapped efficiently  with the stored one
in a reversible process.

We refer to the BSR of Fig.~\ref{fig:circuit_SR}(a),    
where fluxons can arrive on only one input LJJ, 
as a {\em 1-input} BSR, 
to distinguish it from a BSR with separate write and read channels, 
cf.~Fig.~\ref{fig:circuit_2input_SR}.
The circuit dynamics of the 1-input BSR is described by the Lagrangian
\begin{eqnarray}\label{eq:Lagrangian}
 \mathcal{L} &=& \mathcal{L}_l + \mathcal{L}_r + \mathcal{L}_I 
 \,, \\
\mathcal{L}_l 
 &=& \frac{E_0 a}{\lambda_J}  \sum_{n \geq 1} \left[
 \frac{1}{2} \frac{(\dot{\phi}_n^{(l)})^2}{\omega_J^2}
 + \cos\phi_n^{(l)}
 - \frac{(\phi_{n-1}^{(l)}-\phi_{n}^{(l)})^2}{2 (a/\lambda_J)^2} \right] 
 \,, \nonumber \\
\mathcal{L}_r 
 &=& \frac{E_0 a}{\lambda_J}  \sum_{n \geq 1} \left[ 
 \frac{1}{2} \frac{(\dot{\phi}_n^{(r)})^2}{\omega_J^2} 
 + \cos\phi_n^{(r)} 
 - \frac{(\phi_{n}^{(r)}-\phi_{n-1}^{(r)})^2}{2 (a/\lambda_J)^2} \right] 
 \,. \nonumber 
\end{eqnarray}
Herein, $\mathcal{L}_{l,r}$ are the Lagrangian components of the left and right LJJ, 
respectively, and $\mathcal{L}_I$ describes the interface which connects them. 
The JJs in the discrete LJJs have capacitance and critical current of $(C_J, I_c)$, 
and each unit cell of length $a$ has the inductance $L$.
The characteristic time, length, speed and energy scales of the LJJ 
are set by the Josephson plasma frequency 
$\omega_J = 2\pi \nu_J = \sqrt{2\pi I_c/(\Phi_0 C_J)}$, 
the Josephson penetration depth, 
$\lambda_J = a \sqrt{\Phi_0/(2\pi L I_c)}$,
the Swihart velocity $c=\omega_J \lambda_J$,
and the energy scale $E_0 = I_c \Phi_0 \lambda_J/(2\pi a)$
(A static fluxon in the LJJ has energy $8 E_0$, cf.~\Eq{eq:Efluxon}).

In our design the inductance in the interface cell is assumed to be 
negligible, as indicated in Fig.~\ref{fig:circuit_SR}(a), 
and in this situation the phase of the rail JJ of the interface is not
independent but fixed by 
\begin{eqnarray}\label{eq:phiBB}
&&\phiBB =
\phiL-\phiR \,, \\
\label{eq:phiL_phiR}
&& \phiL := \phi_{n=0}^{(l)} \quad\text{and}\quad
 \phiR := \phi_{n=0}^{(r)}
 \;,
\end{eqnarray}
where we introduce shorthand notations for the termination JJ phases in 
\Eq{eq:phiL_phiR}.
With this approximation, the interface Lagrangian corresponding 
to Fig.~\ref{fig:circuit_SR}(a) reads
\begin{eqnarray}\label{eq:Lc_BB1_Lshunted}
&&\mathcal{L}_I 
 = \frac{E_0 a}{\lambda_J} \left\{ 
 \frac{1}{2} \frac{\CJab}{C_J \omega_J^2} \left[\dot{\phi}_L^2 + \dot{\phi}_R^2 \right] 
 + \frac{1}{2} \frac{\CJbb}{C_J} \frac{(\dot{\phi}_L - \dot{\phi}_R)^2}{\omega_J^2} 
 \right. \nonumber \\
&&\hspace*{1.5cm}
 + \frac{\IJab}{I_c} \left[ \cos\phiL + \cos\phiR \right] 
 + \frac{\IJbb}{I_c} \cos(\phiL-\phiR)  \nonumber \\
&&\hspace*{1.5cm} \Biggl.
 - \frac{1}{2} \frac{L \lambda_J^2}{L_s a^2} (\phiL - \phiR + 2\pi f_E)^2
 \Biggr\}  
\end{eqnarray}
where the parameter $f_E$ quantifies an external flux $f_E \Phi_0$
applied to the storage cell, cf.~Fig.~\ref{fig:circuit_SR}(a).
A finite $f_E$ may be useful during the initialization of the BSR, 
i.e. the initial loading of an SFQ into the storage cell, 
but it is not necessary in principle. 
During regular BSR operations, an SFQ is already 
stored and $f_E$ is set to zero. 
In this work we present results on: regular BSR operations (where $f_E=0$)
and initialization results using $f_E=0$.

\subsection{Steady states of the circuit (before input)}\label{sec:SR_initialization}

To analyze the bit-storage characteristics of the BSR
we first study the steady states of the BSR circuit
(in the absence of a moving fluxon). 
It is helpful to first compare the BSR to the earlier ballistic gate circuit 
without a storage cell. 
Schematically, in the limit of infinite storage cell inductance $L_s \to \infty$ 
the BSR circuit is equivalent to the circuit of the ID and NOT gates \cite{WusOsb2020_RFL}. 
The steady states of \Eq{eq:Lagrangian}
are then given by uniform phase fields in the left and right LJJ, 
$\phi^{(l)}_n = 2\pi k_L$ and 
$\phi^{(r)}_n = 2\pi k_R$ ($k_{L,R} \in \mathbb{Z}$), 
while the rail phase assumes the value $\phiBB = 2\pi (k_L-k_R)$.
Herein, the integers $k_{L,R}$ label the `vacuum' states (ground states) 
of the $\phi$-periodic LJJ potential \cite{Rajaraman}. 
Similar to uncoupled LJJs, all configurations $(k_L,k_R)$ are degenerate here,
and the dynamics are not dependent on their initial values.  
When a fluxon from the left LJJ is scattered forward 
without (with) polarity inversion it realizes an ID (NOT) gate; 
it transfers the system from a state with $(k_L,k_R)$ 
to the state with $(k_L + 2\pi \sigma, k_R \pm 2\pi \sigma)$. 
By the way, the dynamics of the NOT gate, but not the ID gate, 
will be used below for the BSR. 

In the presence of finite $L_s$ in the BSR
the degeneracy of different configurations
$(k_L,k_R)$ is lifted due to the contribution 
$\propto (\phiL - \phiR)^2/(2L_s)$ in the potential, cf.~\Eq{eq:Lc_BB1_Lshunted}.
Large values of the rail phase $\phiBB=\phiL-\phiR$ 
(and of the vacuum level difference $2\pi(k_L-k_R)$ 
to the left and right of the interface) become energetically inaccessible. 
At finite $|k_L-k_R| > 0$, 
while the LJJ phases far away from the interface 
are still confined to their respective vacuum levels $2\pi k_{L,R}$, 
the LJJ phases near the interface are perturbed.
We therefore model the LJJ phases (in the absence of a fluxon)
as bound states with evanescent fields of the form
\begin{eqnarray}\label{eq:boundstate_kLR}
 \phi_n^{(l)} &=& (\phiL - 2\pi k_L) e^{-\mu a n} + 2\pi k_L \\
 \phi_n^{(r)} &=& (\phiR - 2\pi k_R) e^{-\mu a n} + 2\pi k_R \,,\nonumber 
\end{eqnarray}
where $\mu$ is the inverse decay length. 
Assuming that the bound-state amplitudes $\phi_{L,R} - 2\pi k_{L,R}$ are small, 
the corresponding rail-phase, $\phiBB=\phiL-\phiR$, 
is approximated by the vacuum level difference, $\phiBB \approx 2\pi (k_L-k_R) = 2\pi S$.
The flux state $S$ in the storage cell
is determined by the difference in configuration at the left and right side of the interface, 
$S=k_L-k_R$.

\begin{figure}[tb]
\includegraphics[width=8.8cm]{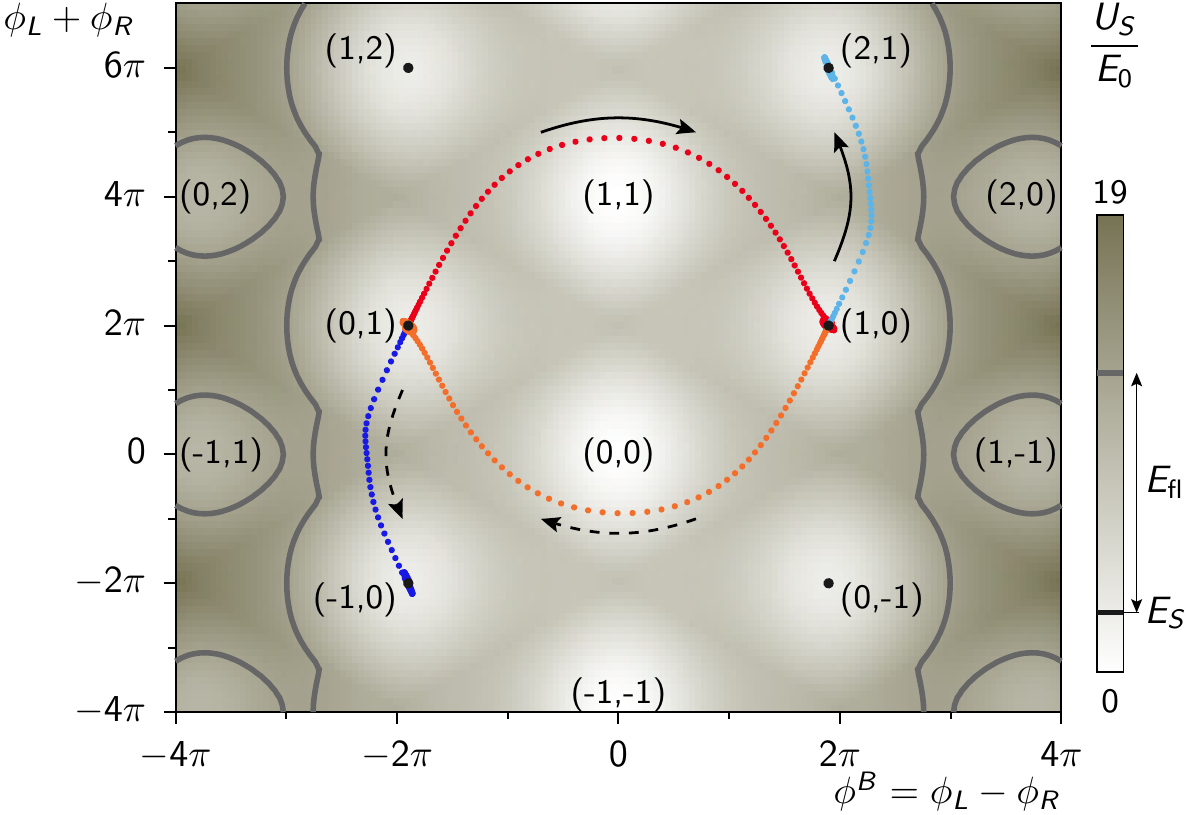}
\caption{
BSR-circuit potential
$U_{S} := \text{min}_{(k_L,k_R)} \left( \tilde U_{(k_L,k_R)} \right)$
in the bound-state approximation, \Eq{eq:boundstate_kLR}, 
as a function of the termination-JJ phases $\phi_{L,R}$: 
The potential is $4\pi$-periodic in the phase sum $\phiL+\phiR$, 
and has an approximate parabolic dependence on
$\phiL-\phiR =\phiBB$ due to energy storage in $L_s$.
Each configuration $(k_L,k_R)$ 
determines a diamond-shaped domain. For $\phiBB \lesssim 4\pi$ 
the diamonds contain a well which supports a stored flux state $S=k_L-k_R$. 
Steady states in wells with $S=\pm 1$ (black points) have
energy $E_S = 2.5 E_0 \approx 2\pi^2 L\lambda_J/(L_s a) S^2 E_0$. 
Additional equipotential lines are shown at $E_S + \Efl$, 
for fluxon energy $\Efl = 10 E_0$, indicating the $\phi_{L,R}$-range 
accessible for an incident fluxon with velocity $v=0.6c$. 
Four trajectories,
$\phiL(t)=\phi_{0}^{(l)}(t)$, $\phiR(t)=\phi_{0}^{(r)}(t)$, are shown (red, blue, orange, light blue points) for different cases of stored SFQ polarity $S = \pm 1$ and incident fluxon. 
These are the termination-JJ phases
obtained from the full circuit simulations.
Solid (dashed) arrows indicate the resulting transitions to another well, 
for $\sigma=1$ ($\sigma=-1$). 
The flux state in the new well is $S'=\pm S$ if $S=\pm \sigma$. 
The system parameters 
are dimensioned such that the potential $U_{S}$ allows 
these transitions, and also the transition from $S=k_L-k_R=0$ 
to $|S|=1$ for the initialization (SFQ-loading) of the BSR.
Note that $U_{S}$ assumes the LJJ fields of the bound-state form,
\Eq{eq:boundstate_kLR}, in the absence of a fluxon.
The superimposed scattering trajectories are therefore not described
by $U_{S}$ alone.
The BSR parameters and parameter ranges are given on the left side of Table \ref{tab:margins}.
} 
\label{fig:U_bs}
\end{figure}

Inserting \Eq{eq:boundstate_kLR} 
in the Lagrangian~\eqref{eq:Lagrangian},
the potential can be expressed, in the limit of small bound-state amplitudes, 
as
\begin{eqnarray}\label{eq:U_bs}
&&\tilde U_{(k_L,k_R)} = \frac{E_0 a}{\lambda_J} \left\{ 
 -\frac{\IJab + I_{c,\text{eff}}}{I_c} \left[ \cos\phiL + \cos\phiR \right] 
  \right. \nonumber \\
&&\hspace*{0.4cm} - \frac{\IJbb}{I_c} \cos(\phiL-\phiR) 
+ \frac{1}{2} \frac{L \lambda_J^2}{L_s a^2} (\phiL - \phiR + 2\pi f_E)^2
\nonumber \\
&&\hspace*{0.4cm}
+\frac{1}{2} \frac{L \lambda_J^2}{L_{\text{eff}} a^2}   
\left[ (\phiL - 2\pi k_L)^2 + (\phiR - 2\pi k_R)^2 \right]  \Biggr\} 
\,.
\end{eqnarray}
Referenced from the interface, each LJJ contribution is 
reduced to an effective JJ and an effective inductance,
\begin{eqnarray}
 I_{c,\text{eff}} &=& I_c / (e^{2\mu a} - 1) \\
 L_{\text{eff}}   &=& L (e^{\mu a} + 1) / (e^{\mu a} - 1) 
 \,,
\end{eqnarray}
where both are in parallel with the corresponding termination JJ ($\IJab$).
In these expressions the inverse decay length $\mu$ of the bound state
is not yet determined. However, we can estimate $\mu$ from the condition that 
the bound state fulfills the dispersion relation in the LJJ bulk, 
$\omega_{\text{bulk}}^2 = \omega_J^2 + 2c^2/a^2 \left(1 - \cosh(a\mu)\right)$. 
Being interested in steady states of the interface, 
we can set $\omega=0$ and obtain the estimate 
$\mu = a^{-1} \cosh^{-1}\left( 1 + a^2/(2\lambda_J^2) \right)$.

The potential shown in Fig.~\ref{fig:U_bs} is obtained from \Eq{eq:U_bs}
by choosing the energy-minimizing configuration, 
$U_{S} := \text{min}_{(k_L,k_R)}\left( \tilde U_{(k_L,k_R)} \right)$,
for each point $(\phi_L,\phi_R)$.
The resulting diamond-shaped domains are labeled in Fig.~\ref{fig:U_bs}
by the locally minimizing $(k_L,k_R)$.
The potential is $4\pi$-periodic in the sum of the phases $\phiL+\phiR$ 
at constant phase difference $\phiL-\phiR$, 
due to the combined $2\pi$-periodicity in the two components.
However, the potential has an approximate parabolic dependence on
$\phiBB = \phiL-\phiR$.
For not too large $\phiBB$, the potential $U_{S}$
has a local minimum in each of the domains $(k_L,k_R)$,
and these steady states correspond to a stored flux state $S=k_L-k_R$. 
Degenerate global minima are found 
at $\phiBB=\phiL-\phiR=0$ in the domains 
with zero stored flux, $(k_L-k_R)=0$.
States with a single stored flux quantum
are found in the domains with $|k_L-k_R|=1$, 
with the local minima at
$\phiBB=\phiL-\phiR \approx 2\pi (k_L-k_R) = \pm 2\pi$.
From the vertical position of the minima, $\phiL+\phiR = 2\pi(k_L+k_R)$, 
it follows that the bound-state amplitudes on the left and right sides of the interface 
are equal and opposite, $(\phiL-2\pi k_L) + (\phiR-2\pi k_R) = 0$.

During normal operation of the BSR, the parameters satisfy 
$f_E=0$,  $\max(\IJbb, \IJab, I_{c,\text{eff}}) \gg \Phi_0/(2\pi L_s)$, 
and $|S| \leq 1$, where 
the bound-state amplitudes are small and the phase fields in the left and right LJJ 
are nearly uniform. 
For a single stored flux quantum $S$ in the BSR, 
we can thus approximate $(\phiL,\phiR)$ in \Eq{eq:boundstate_kLR} as $(2\pi k_L, 2\pi k_R)$. 
From \Eq{eq:boundstate_kLR} it follows that the stored energy relative to the empty BSR ($S=0$) is
\begin{eqnarray}\label{eq:E_S__bs__ccm}
 E_S \approx \frac{2\pi^2 L\lambda_J}{L_s a} (k_L - k_R)^2 E_0 
 = \frac{2\pi^2 L\lambda_J}{L_s a} S^2 E_0 
 \,.
\end{eqnarray}
If the BSR is initialized with $|S| \leq 1$, 
an incoming fluxon with velocity $v$ and energy
\begin{equation}\label{eq:Efluxon}
\Efl(v) = 8 E_0\left(1-(v/c)^2\right)^{-1/2} 
\end{equation}
can transfer the BSR into a new stored flux state  
with energy $E_S' \leq E_S + \Efl(v)$. 
The equipotential lines in Fig.~\ref{fig:U_bs} indicate the corresponding
$\phi_{L,R}$-range accessible from $|S|=1$.
Energetically, it is not possible to load a 2nd SFQ ($|S|=2$) 
into the storage cell,
whereas transitions to other states with $|S|=1$ or $S=0$ are energetically possible.

\subsection{Fluxon scattering dynamics}\label{sec:SR_operation}

\begin{figure}[tb]\centering
\includegraphics[width=8.8cm]{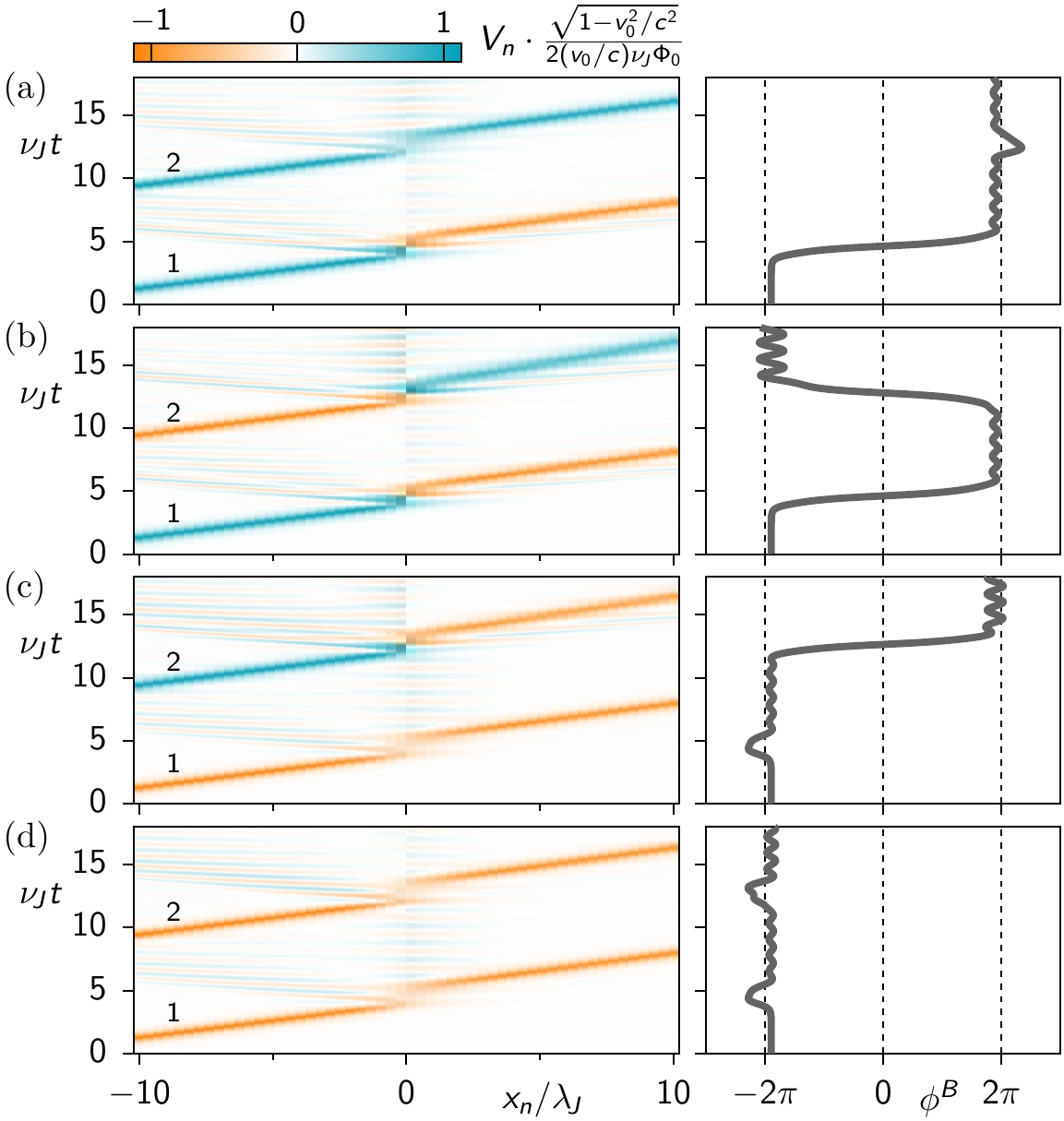}
\caption{
Simulated operations of the shift register initialized with the stored flux quantum $S=-1$,
under four different input sequences of two fluxons: 
$(\sigma_{1}, \sigma_{2}) = (1,1), (1,-1), (-1,1), (-1,-1)$. 
Left panels show dynamics of JJ-voltages $V_n$ at positions 
$x_n \lessgtr 0$ in the left (input) and right (output) LJJ, respectively. 
The color scale shows blue tracks for fluxons $(\sigma=1)$ 
and orange for antifluxons $(\sigma=-1)$. 
Right panels show evolution of the rail-JJ phase $\phiBB$ 
from initial state $\phiBB \approx 2\pi S$.
In cases where $\sigma = -S$, 
the fluxon scatters forward as an output fluxon with inverted polarity, 
$\sigma' = S$, 
and the stored state after the scattering becomes $S'=\sigma$. 
In the other cases, $\sigma = S$, the fluxon is simply transmitted with short delay, and the stored state remains unchanged. 
All cases fulfill ($S',\sigma'$) = SWAP($S,\sigma$),
thus generating the state map of a 1-bit shift register, cf.~Fig.~\ref{fig:circuit_SR}(b).
The BSR parameters, as listed in Table \ref{tab:margins}, 
are the same as in Fig.~\ref{fig:U_bs}
and input fluxons enter with $v_0 = 0.6 c$.
}
 \label{fig:phi_x_t_SR_190619__2fluxonevent}
\end{figure}

Figure~\ref{fig:phi_x_t_SR_190619__2fluxonevent} illustrates 
the BSR operation, where each of the four subfigures shows the circuit
simulations for two consecutive input fluxons. 
In all four cases, the BSR is assumed to initially contain 
a stored flux quantum $S=-1$.
This means that the circuit is
initialized in a bound state of the form of \Eq{eq:boundstate_kLR},
with $(k_L,k_R)=(0,1)$ and corresponding steady state values of $\phi_{L,R}$.
The incoming fluxon(s) are treated in simulation as additional contributions 
to the initial phase and voltage distribution in the left LJJ,
far away from the interface. 
An input fluxon (antifluxon), 
which has positive (negative) polarity $\sigma=1 (-1)$, 
is parametrized by the ideal phase distribution
$\phi(x,t) = 4 \arctan\left(\exp\left(-\sigma (x - v t)/W) \right)\right)$
with velocity $v$ and width $W = \lambda_J (1-v^2/c^2)^{1/2}$.
This corresponds to a positive (negative) voltage pulse with maximum (minimum)
$\pm 2 \Phi_0 \nu_J (v/c) (1-v^2/c^2)^{-1/2}$.
We compute the fluxon dynamics from numerical integration of 
the $(N_l + N_r + 3)$ classical circuit equations of motion 
for the $(N_l + N_r)$ JJs in the LJJs,
together with the termination and rail JJs of the interface. 
The left panels of Fig.~\ref{fig:phi_x_t_SR_190619__2fluxonevent}
show the resulting JJ-voltages $V_n^{(l,r)}$ 
at positions $x_n =  \mp a (n + 1/2)  \lessgtr 0$ ($n=0,1,2,\ldots$)
in the left and right LJJ. 
The position of the interface is $x = 0$.
The right panels of Fig.~\ref{fig:phi_x_t_SR_190619__2fluxonevent} show 
the evolution of the rail phase $\phiBB(t)$
from the initial value $\phiBB(0) \approx -2\pi$.

Figure \ref{fig:phi_x_t_SR_190619__2fluxonevent}(a) shows at the earliest times, 
a first fluxon with polarity $\sigma=+1$ traveling 
in the left LJJ with nearly constant speed $v_0 = 0.6 c$.
As it reaches the interface, the fluxon breaks into two parts,
i.e.~its phase- and voltage-fields become discontinuous. 
In the process, the energy of the fluxon is coherently transferred 
to a localized excitation involving
the left and right LJJs in form of time-dependent evanescent fields. 
The localized excitation lasts long enough for its own brief oscillation, 
and afterwards generates a large field profile in the right LJJ, 
which eventually moves as a free fluxon in the right LJJ, 
away from the influence of the interface.
During the whole process, the phase $\phiBB$ of the rail JJ changes 
monotonously from $\phiBB \approx -2\pi$ to $\approx 2\pi$. 
This $4\pi$-phase change indicates the simultaneous transfer 
of the input fluxon's (positive) SFQ-state $\sigma=1$ into the storage cell 
and of the initially stored (negative) SFQ $S=-1$ out of the storage cell. 
Thusly after the scattering, the new orientation of the stored flux quantum is $S'=1$, 
and the output fluxon carries the negative SFQ-state, $\sigma'=-1$, 
as indicated by the negative sign of the voltage peak.
The fluxon scattering dynamics in this case are similar to that of 
the fundamental (1-bit) NOT gate \cite{WusOsb2020_RFL} (without a storage cell).

When the second input fluxon with $\sigma=1$ arrives
(Fig.~\ref{fig:phi_x_t_SR_190619__2fluxonevent}(a), $\nu_J t \approx 12$), 
the input SFQ state now equals the stored SFQ state, $S=1$. 
The resulting scattering dynamics at the interface therefore differs 
significantly from that of the preceding fluxon. 
The interface here acts mostly as a low potential barrier by which the fluxon is slowed down 
temporarily while retaining its fluxon identity, with unchanged polarity. 
During the fluxon transmission the rail JJ is only weakly excited away from 
$\phiBB \approx 2\pi$, indicating that no significant flux transfer occurs.  
Accordingly, both the fluxon's state $\sigma$ and the stored flux state $S$
are unchanged in this process. 
While the result of the fluxon scattering is here the same as in the fundamental ballistic ID gate
(the polarity of the outgoing fluxon is identical to that of the incoming one), 
the scattering dynamics is different: 
the fluxon here retains its topological identity throughout the process,
whereas in the fundamental ID gate it breaks up into two partial fluxons at the interface 
and generates a large localized oscillation as a result 
(which has a longer duration compared with the temporary oscillation 
in the fundamental NOT gate). 
The difference of the transmission-type dynamics of the BSR
compared with the dynamics of an actual ID-gate originates from the added term 
$(\phiBB)^2/(2 L_s)$ in the interface potential, \Eq{eq:Lc_BB1_Lshunted}. 
It limits the $\phiBB$-range accessible with the fluxon's initial energy $\Efl$
(see Fig.~\ref{fig:U_bs} and discussion in Sec.~\ref{sec:SR_initialization}).
As a result, the rail phase in the BSR cannot increase from an initial value 
$\phiBB \approx 2\pi$ by $\Delta \phiBB \approx +4\pi$.
The transmission-type dynamics creates an advantage for the margins of the BSR 
(see Sec.~\ref{sec:margins})
relative to the fundamental ID gate, which has somewhat sensitive margins 
in comparison with the fundamental NOT gate due to the longer resonant oscillation.
In summary, the inductor $L_s$ enables bit storage, 
changes the dynamics relative to previous RFL gates, 
and improves operation margins compared to an earlier gate.

Figure \ref{fig:phi_x_t_SR_190619__2fluxonevent} demonstrates
that the scattering dynamics results for all state pairs ($S,\sigma$)
in a new state pair, 
which is related to the old one in the form of a SWAP-operation, 
($S',\sigma'$) = SWAP($S,\sigma$). 
The ballistic scattering dynamics in the BSR circuit thus generates
the state map of a 1-bit shift register, Fig.~\ref{fig:circuit_SR}(b).
As described above, two different types of scattering dynamics are involved, 
and in both types the SWAP happens with almost ideal efficiency
(cf.~Sec.~\ref{sec:margins}).

The BSR operations can be illustrated as inter-well transitions 
in the circuit potential $U_{S}$, 
induced by the incoming fluxon.
This is illustrated in Fig.~\ref{fig:U_bs} by the 
trajectories $(\phiL,\phiR)(t)$ (data points), 
where $\phi_{L,R}(t)$ are taken from the full circuit simulation.
Note that the circuit potential shown in Fig.~\ref{fig:U_bs} 
assumes the LJJ fields have the form of a bound state, \Eq{eq:boundstate_kLR}, 
but does not take into account the fluxon.
For example, a fluxon with $\sigma=1$ leads to a transition (red) 
from the initial state in the well with $(k_L,k_R)=(0,1)$ 
to the well with $(k_L,k_R)=(1,0)$, 
and this changes the stored flux state $S=-1 \to 1$.
In contrast, a fluxon with $\sigma=-1$ induces an $S$-preserving transition (blue), 
namely to the well with $(k_L,k_R)=(-1,0)$, which is fully equivalent 
to the initial well. 
The underlying circuit dynamics for these two processes corresponds to the first 
fluxon scattering in Figs.~\ref{fig:phi_x_t_SR_190619__2fluxonevent}(a) and (c), 
respectively.
Equivalent dynamics are observed 
for an initial stored flux state $S=1$, 
e.g. initially $(k_L,k_R)=(1,0)$, 
if the polarity of the incoming fluxon is inverted at the same time.
Thus, an incoming fluxon with $\sigma=-1$ ($\sigma=1$) induces a transition 
that inverts (preserves) $S$, as shown by the orange (light blue) trajectories.

The dynamics shown in Fig.~\ref{fig:phi_x_t_SR_190619__2fluxonevent} 
illustrate the regular BSR operations, 
where initially a flux quantum is already stored in the storage cell. 
Without this initialization, the BSR will not perform all the intended 
reversible operations.
To initialize a BSR, an SFQ can be loaded into the empty storage cell
by sending in a fluxon on the bit line. 
As the interface cell is designed with minimal inductance, it cannot 
hold an SFQ. Therefore, once the fluxon is stopped near the interface, 
its flux is transferred into the storage cell.
We find that it is possible to load the BSR in this way with an SFQ, 
using no external flux ($f_E=0$) and a fluxon of nominal energy 
$\Efl = 10 E_0$. 
A lower-energy fluxon ($\Efl < 10 E_0$) might be better for loading 
because less excess energy would have to dissipate 
prior to reaching a quiescent state, but due to a potential barrier 
a very slow fluxon reflects from the interface instead of trapping. 
We found that the potential barrier can be lowered by applying an external flux 
($f_E \approx 0.25$), 
and in that case a low energy fluxon ($\Efl \approx 8.2 E_0$) 
is successfully loaded into the storage cell.

The above initialization procedure seems suitable for individual BSRs. 
Another procedure may be favorable in a large circuit with many BSR gates. 
A suitably designed circuit could in principle be made to trap flux solely 
in the BSR storage cells. To initialize many BSR gates in such a circuit 
one would cool through the superconducting transition in a magnetic field, 
and then turn off the field.

\subsection{Sequentially arranged shift registers}\label{sec:serialSR}

\begin{figure}[tb]
\includegraphics[width=8.8cm]{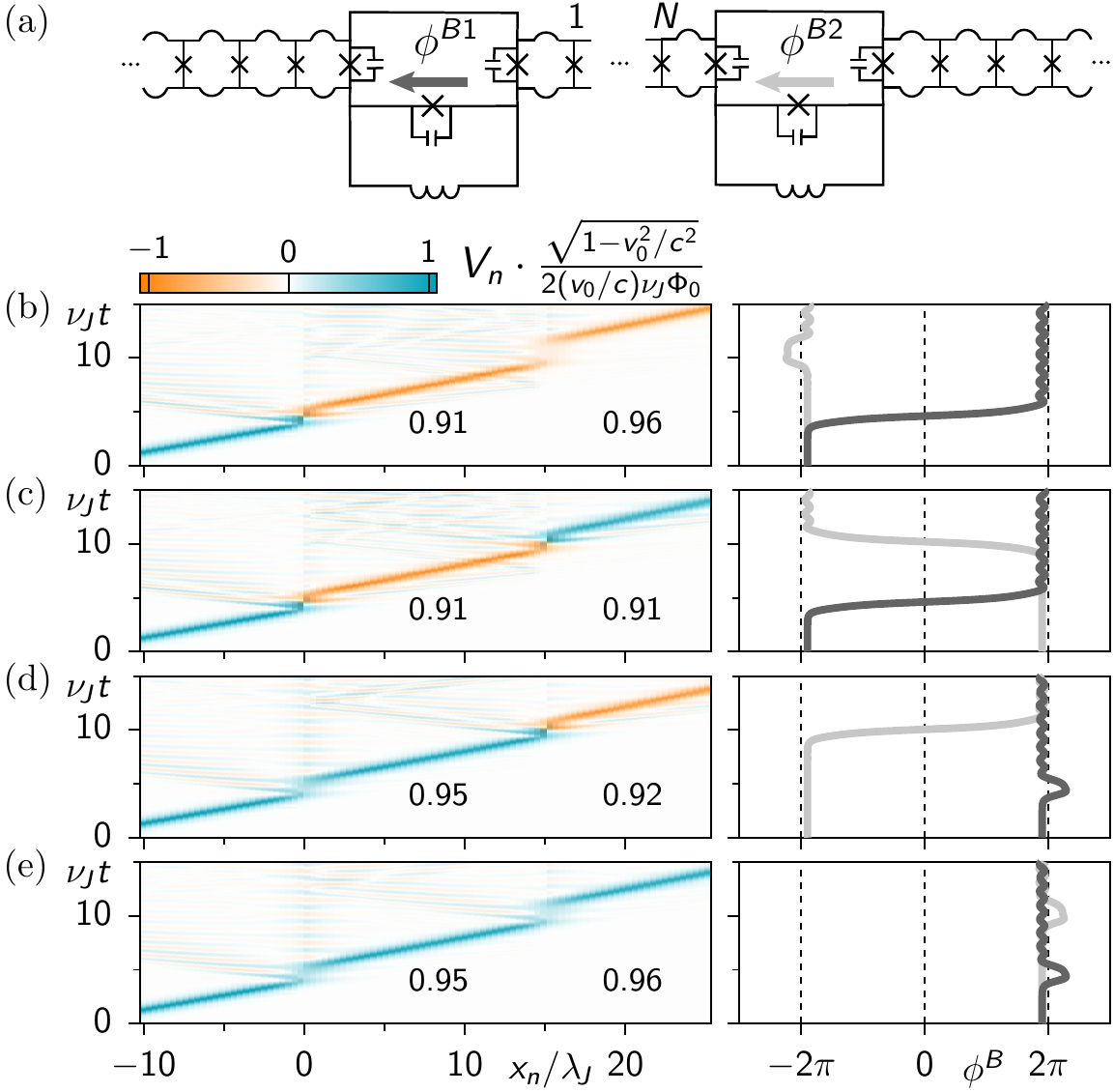}
\caption{
A 2-bit Serial-In-Serial-Out (SISO) shift register (a) and its dynamics (b-e) 
for a single input fluxon, $\sigma = +1$, 
and four different initial configurations of the two stored bits:
$(S_1,S_2)=(-1,-1)$, $(-1,1)$, $(1,-1)$, and $(1,1)$.
As in Fig.~\ref{fig:phi_x_t_SR_190619__2fluxonevent},
the left panels show the JJ voltages 
$V_n$ in the LJJs, where a fluxon ($\sigma=1$) is seen as a blue track 
and an antifluxon ($\sigma=-1$) is seen as an orange track. 
The two BSR are located at $x_n=0$ and at 
$x_n = (N+2)a = 40 a \sim 15 \lambda_J$, respectively, 
though smaller distances are also possible.
The right panels show the evolution of the rail-JJ phases $\phi^{Bi}$ of the two BSR ($i=1,2$). 
The numbers printed in each of the left panels
are the relative speed $v_1/v_0$ ($v_2/v_1$) after passing through the first (second) 
BSR.
The BSR parameters are the same as in 
Fig.~\ref{fig:phi_x_t_SR_190619__2fluxonevent}.
}
\label{fig:serial2bit_SR}
\end{figure}

A multi-bit shift register can be constructed from sequentially arranged BSRs,
constituting a Serial-In-Serial-Out (SISO) register. 
As an example, Fig.~\ref{fig:serial2bit_SR} shows 
a 2-bit serial shift register and its dynamics.
The two bits are stored in one of four different configurations 
$(S_1,S_2)=(-1,-1)$, $(-1,1)$, $(1,-1)$, and $(1,1)$.
Subfigures~\ref{fig:serial2bit_SR}(b-e) show the gate dynamics
for each of these initial configurations
and a single input fluxon, here with $\sigma=+1$.
The two BSR are located at $x=0$ and $x\approx 15 \lambda_J$ 
(separated by $N+2 =40$ LJJ cells). 
As in Fig.~\ref{fig:phi_x_t_SR_190619__2fluxonevent},
the stored flux quanta can be inferred from the values
of the rail-JJ phases $\phi^{B1}$ and $\phi^{B2}$ in the right panels of 
each subfigure, using that $\phi^{Bi} \approx 2\pi S_i$ ($i=1,2$). 
The operation of the entire 2-bit shift register is powered by the 
energy of the input fluxon, which looses only a fraction of its kinetic energy 
in each of the scatterings. 
The numbers printed in the left panels are the output-to-input velocity ratios
after each scattering, $v_{1}/v_{0}$ and $v_{2}/v_{1}$,
where again we use initial velocity $v_0=0.6c$,
corresponding to an initial energy of $\Efl(v_0) = 10 E_0$.
The lowest velocity ratio ($0.91$) corresponds to $95\%$ energy conservation
according to \Eq{eq:Efluxon}.

For each scattering type (NOT and transmission), 
both stages of the 2-bit shift register give approximately 
the same velocity ratios ($\approx 0.91$ for NOT, $\approx 0.95$ for transmission),
corresponding to those of a 1-bit BSR (see Sec.~\ref{sec:margins}).
The observed small variability in forward-scattering efficiency 
at the 1st and 2nd BSR 
(e.g. between 0.95 and 0.96 for the two consecutive transmissions in panel (e))
can be attributed to the presence of fluctuations in the connecting LJJ 
and at the 2nd BSR prior to the fluxon arrival there. 
These fluctuations are emitted from the first BSR during the first scattering event. 

Due to a relatively sharp lower cut-off velocity of the 
transmission dynamics of one BSR, cf.~Fig.~\ref{fig:margins}(f), 
we believe that $k=2$ may be the maximum number of sequentially arranged BSRs  
for a realistic circuit design without added power. 
It is beyond the scope of the paper to discuss how to reach a $k$-bit 
sequential shift register with larger $k$, but we plan to address this in future work.

\subsection{2-input shift register}\label{sec:2-inputSR}

\begin{figure}[b]\centering
 \includegraphics[width=7cm]{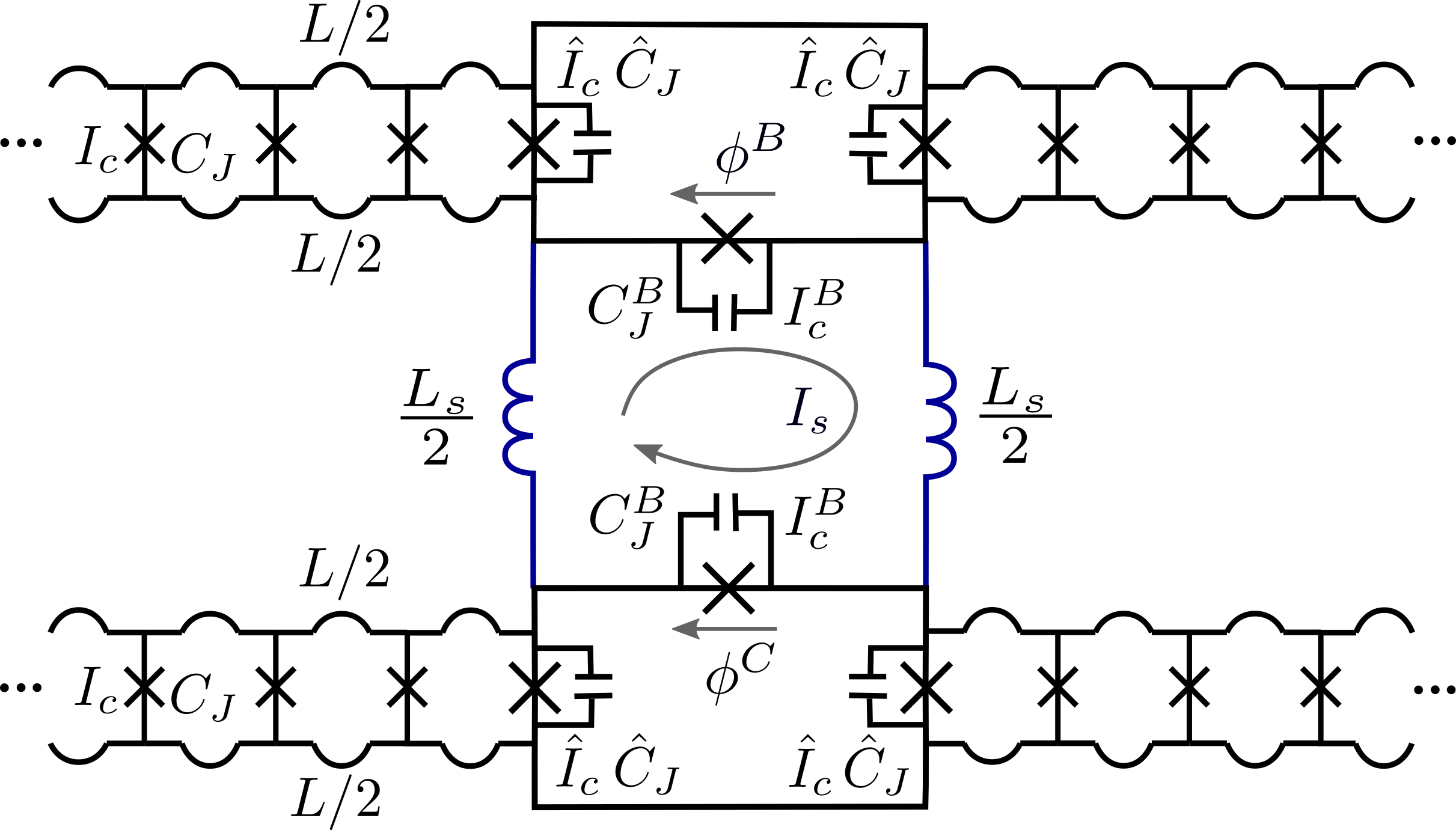}
\caption{ 
 Circuit schematic of the 2-input shift register, 
 where the upper and lower LJJ pairs form separate fluxon scattering channels
 (bit lines)
 with a shared storage cell between them.
 The inductance of the storage cell is $L_s$ 
 and the interface cells of both scattering channels are symmetric.
 Efficient BSR operation takes place with parameter values given in 
 Fig.~\ref{fig:phi_x_t_SR_190619__2fluxonevent},
 and operation margins given in the right side of Table \ref{tab:margins}.
 }
\label{fig:circuit_2input_SR}
\end{figure}

The 1-bit BSR, shown in Fig~\ref{fig:circuit_SR}(a), has one input LJJ
and one output LJJ. 
Together they form a single channel called a bit line for the forward-scattering fluxon.
This {\it 1-input} 1-bit BSR may be generalized to 
a {\it 2-input} 1-bit BSR, where a storage cell
is shared between two such scattering channels, 
as shown in Fig.~\ref{fig:circuit_2input_SR}. 
We have verified in simulations that the 2-input structure
also acts as an energy-efficient 1-bit BSR,
using the same interface parameters as for the 1-input 1-bit BSR, 
cf.~Table \ref{tab:margins}. 
For the operation of this BSR, it is irrelevant which of the 
two input LJJs a fluxon is sent in on -- 
the dynamics for both input cases is equivalent
and it is qualitatively equivalent to the dynamics of the 1-input BSR. 
In the 2-input device the role of the rail-JJ phase $\phiBB$ of the 1-input BSR 
is taken over by the phase difference of the two rail JJs, 
$\phiBB \to \phiBB-\phiCC$. 
Motivated by the small difference in parameter margins 
between the 1-input and the 2-input version of the BSR, 
cf.~Table \ref{tab:margins}, 
we expect that a version with many inputs would also operate. 
This implies that a stored bit of information could be routed to one of many outputs. 

Despite the shared storage cell
there is no strong dynamic coupling between the upper and lower part of the gate.
By that we mean that even during the NOT-type scattering
an input fluxon on the upper (lower) LJJ induces a large phase 
change of $4\pi$ only in the adjacent $\phiBB$ ($\phiCC$), 
and a relatively small temporary excitation in $\phiCC$ ($\phiBB$),
resulting in an output fluxon only on the upper (lower) output LJJ. 
A similar observation holds for the transmission-type scattering, 
however in this case no significant phase change takes place even in the adjacent rail JJ. 
With this property, the 2-input BSR may be used as an SFQ memory with 
separate write and read lines.

\subsection{Margins}\label{sec:margins}

\begin{figure}
\includegraphics[width=8.8cm]{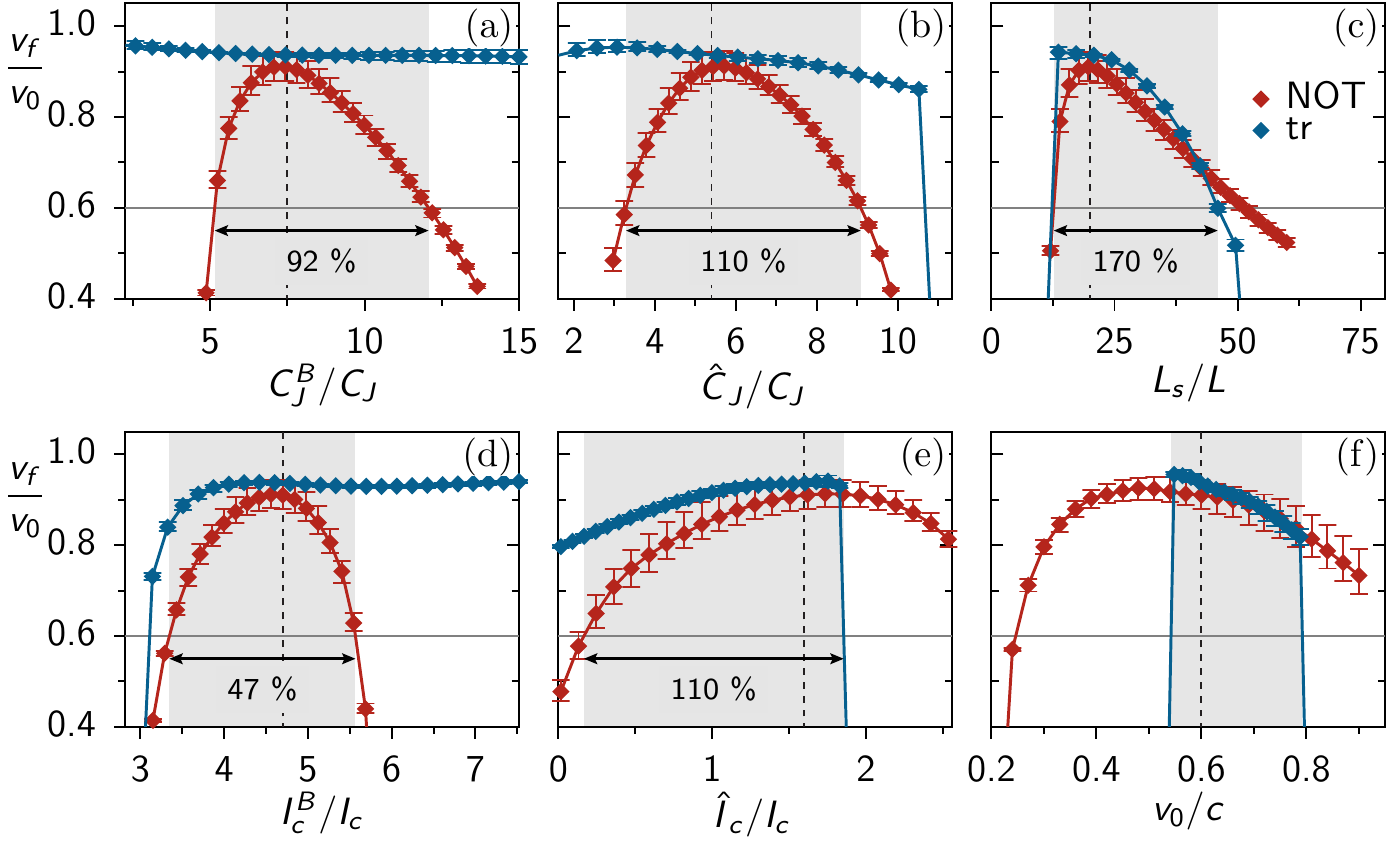}
\caption{
Margins of 1-input BSR:
the output-to-input velocity (retention) ratio, $v_f/v_0$, 
(a-e) for $v_0/c=0.6$, as a function of varied 
interface parameters, $\CJbb$, $\CJab$, $L_s$, $\IJbb$, and $\IJab$, respectively;
(f) for fixed interface parameters but varied initial velocity $v_0$.
In (a-f), all parameters except the varied one
are kept constant at values given in Table \ref{tab:margins}.
Error bars mark the amplitudes of velocity oscillations (an uncertainty) after scattering.
Shaded regions illustrate the ranges wherein both scattering types 
(NOT (red) and transmission (blue))
fulfill $v_f/v_0 \geq 0.6$, i.e. $\Efl(v_f)/\Efl(v_0) \geq 0.86$.
This condition produces the margins given in the left side of Table \ref{tab:margins}.
}
\label{fig:margins}
\end{figure}

\begin{table}[b]
\renewcommand\arraystretch{1.5}
\footnotesize
\begin{tabular}{|l|l||c|c|c||c|c|c|}
\hline
\multicolumn{2}{|c||}{parameter} & \multicolumn{3}{c||}{1-input BSR} 
& \multicolumn{3}{c|}{2-input BSR}\\ 
\hline
$\:p$ & $p_0$ 
& $\negthinspace\frac{p_\text{min} - p_0}{p_0}\negthinspace$ 
& $\negthinspace\frac{p_\text{max} - p_0}{p_0}\negthinspace$ 
& $\frac{\Delta p}{p_0}$ 
& $\negthinspace\frac{p_\text{min} - p_0}{p_0}\negthinspace$ 
& $\negthinspace\frac{p_\text{max} - p_0}{p_0}\negthinspace$ 
& $\frac{\Delta p}{p_0}$ 
\\
\hline 
$\CJbb/C_J$ 
& 7.5 
& -31\% & \phantom{1}+61\% & \phantom{1}92\%   
& -35\% & \phantom{1}+56\% & \phantom{1}91\% \\
$\CJab/C_J$ 
& 5.4 
& -39\% & \phantom{1}+68\% & 107\%  
& -37\% & \phantom{1}+79\% & 116\% \\
$\IJbb/I_c$ 
& 4.7 
& -29\% & \phantom{1}+18\% & \phantom{1}47\%   
& -30\% & \phantom{1}+16\% & \phantom{1}46\% \\
$\IJab/I_c$ 
& 1.6 
& -90\% & \phantom{1}+16\% & 106\%  
& -78\% & \phantom{1}+20\% & \phantom{1}98\% \\
$L_s/L$ 
& 20  
& -37\% & +130\% & 167\%     
& -35\% & +124\% & 159\% \\
\hline
\end{tabular}
\caption{
Interface parameters and margins 
for 1-input BSR, Fig.~\ref{fig:circuit_SR}(a),
and 2-input BSR, Fig~\ref{fig:circuit_2input_SR}.
Margins are defined by a required output-to-input velocity ratio $v_f/v_0 \geq 0.6$, 
cf.~Fig.~\ref{fig:margins}, corresponding to an energy efficiency 
$\Efl(v_f)/\Efl(v_0) \geq 0.86$ for initial $v_0=0.6 c$.
This condition is met by all regular BSR operations, 
i.e.~for any combinations of $S=\pm 1$ and $\sigma = \pm 1$.
In case of the 2-input BSR, it is also independent 
of the choice of input LJJ.
In both BSR types the parameters allow for a velocity retention up to 
$v_f/v_0 = 0.91$, cf.~Fig.~\ref{fig:margins}.
}
\label{tab:margins}
\end{table}

An optimal set of circuit parameters for an energy-efficient BSR
are given by the first and second columns in Table \ref{tab:margins}. 
These parameters optimize the elastic nature of both scattering types 
(NOT and transmission) of the BSR operation, such that the dominant 
fraction of the input fluxon's energy is conserved in the forward-scattered fluxon.
The resulting output-to-input velocity ratio $v_f/v_0$ 
of the optimized dynamics amounts to $0.91$ and $0.95$, respectively.
For an input fluxon with $v_0 = 0.6c$ 
the average energy efficiency of the BSR therefore is 96\% 
according to \Eq{eq:Efluxon}.
Figure \ref{fig:margins} shows the output-to-input velocity ratio under
variations of different parameters. 
Setting the minimum ratio $v_f/v_0$ to $0.6$ 
(corresponding to an energy efficiency 
$\Efl(v_f)/\Efl(v_0) \geq 0.86$ for initial $v_0=0.6 c$), 
we find the operation margins for the BSR, 
as shown in the next three columns in Table \ref{tab:margins}.
The current limiting factor of the BSR design, 
as shown in Fig.~\ref{fig:margins}(f),
is the somewhat restricted range of input velocities for the transmission-type 
BSR dynamics. 
In this case, the interface's potential barriers which are proportional to 
$\IJbb$ and $\IJab$ 
impose a sharp lower operation limit of $v_0 \geq 0.53 c$,
though other parameters can be used to reduce this lower velocity limit. 
Of the parameter margins, $\IJbb$ is the smallest with a range of $47\%$.

In addition to the variation of interface parameters presented in Fig.~\ref{fig:margins}, 
we have studied gate robustness under variation of LJJ parameters. 
The resulting margins are generally favourable too, with the smallest range 
of $66\%$ appearing under variations of the LJJ cell inductances $L$, 
as shown in Appendix \ref{app:CCM}.

The margins of the 2-input BSR, shown in the last three columns of Table \ref{tab:margins}, 
are almost the same as those of the 1-input BSR. 
This is consistent with our observation earlier, 
that the dynamics on each (upper or lower) bit line  
is only very weakly affected by the presence of the other bit line. 
As long as there is no excitation on the extra bit line, 
it mainly acts as an inductance added to the storage loop. 
Therefore, we expect that a BSR gate with more bit lines (an k-input BSR)
will operate similarly well.

\subsection{Asynchronous gate timing}\label{sec:timing}

\begin{figure}
\includegraphics[width=\columnwidth]{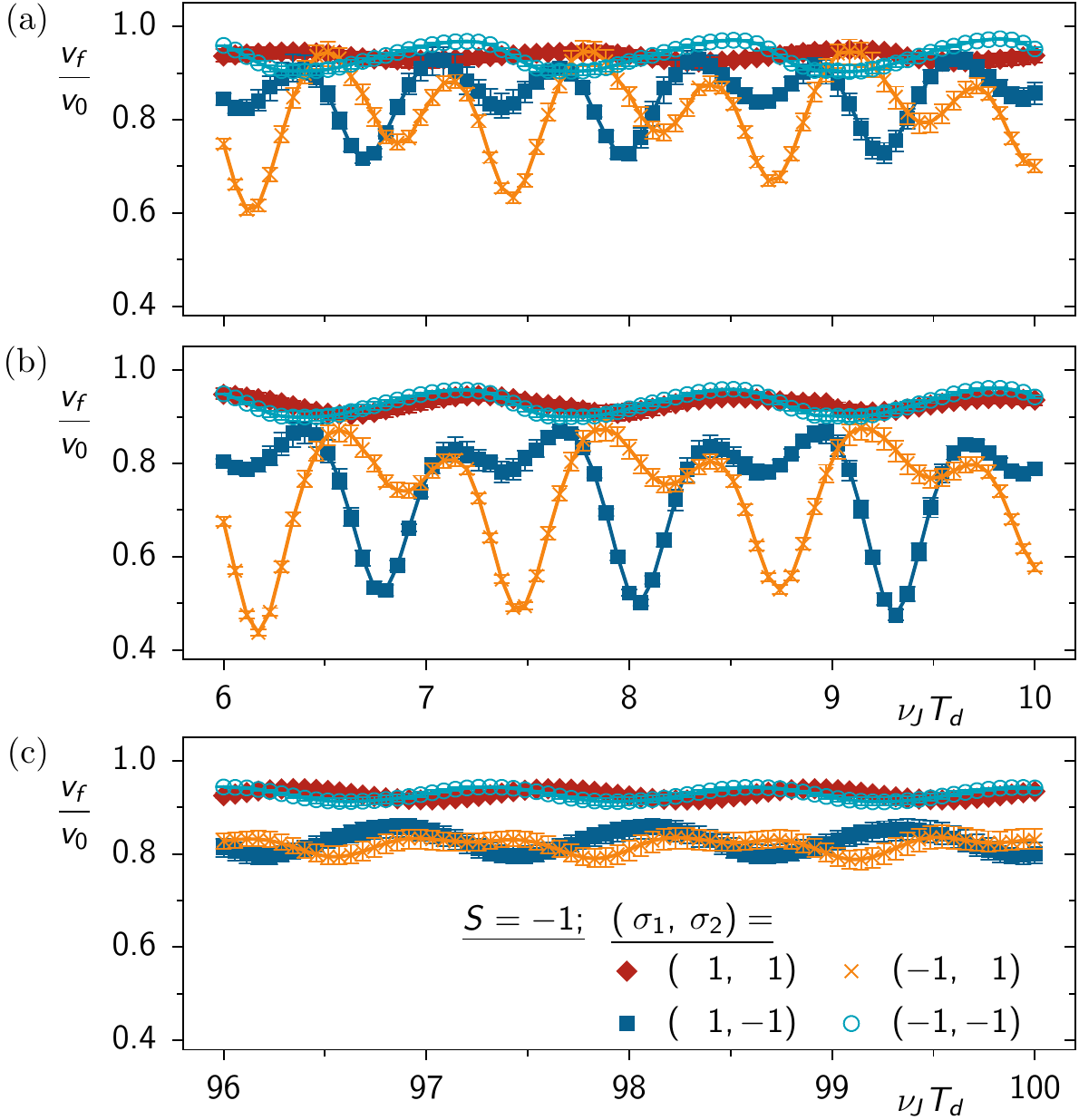}
\caption{
Output-to-input velocity ratio of 1-input BSR
under variation of the delay time $T_d$ of the input fluxon to a preceding input fluxon, 
for (a) undamped and (b,c) weakly damped JJs in the SR interface.
The BSR efficiency depends of course on the gate operation of the current input fluxon (polarity $\sigma_2$), but also on the preceding gate operation (input polarity $\sigma_1$ and SR initialized with $S=-1$), as illustrated by the four different curves in each panel.
The output velocity $v_f$ oscillates with approximate periodicity $1/\nu_J$
due to energy remaining at the SR interface after the preceding gate operation.
The interface parameters are those given in Table \ref{tab:margins}, 
where in (b,c) we have added shunt resistors to the three interface
JJs, corresponding to a 
loss tangent of $\tan\!\delta(\omega=\omega_J) = 4\cdot 10^{-3}$:
$\hat R_J = 1/(\tan\!\delta\, \omega_J \CJab)$
and $R_J^B = 1/(\tan\!\delta\, \omega_J \CJbb)$, 
cf.~Fig.~\ref{fig:circuit_SR}(a).
By adding circuit damping the $v_f$-oscillations increase slightly and the average $v_f$ is slightly reduced, 
see panel (b) relative to (a). 
However, for long delay times, damping will diminish $v_f$-variations, 
see panel (c) relative to (b), and allow asynchronous gate operations
independent of input timings. 
}
\label{fig:variation_Tdelay}
\end{figure}

In most SFQ logic, including RSFQ and its descendants, 
the gates are clocked, and the SFQ data bit is processed in gates once the clock pulse arrives. Although data bits do not need to arrive at a precise time, the gates are still considered synchronous because the data SFQs must be processed with a clock pulse. 

In asynchronous reversible logic gates, the requirements on the arrival time are reduced: operations are not clocked, and bits only need to arrive in a definite order. 
To achieve definite order, the input constraint is a negligible direct interaction between successive input bits. 
In the LJJ the interaction strength decays exponentially for large distances 
compared to $\lambda_J$ \cite{Rubinstein1977}, 
and in our LJJs we find that $10\lambda_J$ is a suitable distance for negligible interactions. 
For an asynchronous 1-input gate like the 1-bit BSR, 
this corresponds to a minimum required delay time between two input fluxons, 
$T_d > \text{max}\left(10\lambda_J/v_0, \tau_{\text{max}}\right)$, 
where $v_0$ is the (common) input velocity and $\tau_{\text{max}}$ 
is the maximum duration of all possible gate operations. 
To explore the role of the delay time $T_d$, this section investigates 
(i) the dependence of the BSR on $T_d$ and (ii) vice versa, 
i.e. the modification of the fluxon delay {\em after} the BSR gate.
Finally, the gate-induced timing uncertainty is compared with that arising from fluxon jitter.

Next we describe uncertainties created by variation of the fluxon delay time 
and different gate operations. 
In the underlying simulations we study a 1-bit BSR initialized with $S=-1$ 
and two consecutive input fluxons with a delay time $T_d$ between them. 
The initialized gate is in a steady state, but the first gate operation 
leaves residual energy at the interface. 
Thus the efficiency of the second operation will depend not only on its own 
dynamics (operation type), but also that of the preceding one. 
As a result, each of the four input combinations $(\sigma_1, \sigma_2)$ 
generate distinct curves (different colors) for the output velocity $v_f$ of the second fluxon,
as shown in Fig.~\ref{fig:variation_Tdelay}. 
These curves exhibit oscillations in $v_f$ with approximate periodicity of $1/\nu_J$, 
due to the constructive or destructive interaction of the input fluxon with the 
oscillations remaining from the preceding gate operation.
While panel (a) shows the situation in absence of damping, 
in panels (b,c) we show the results for damping added to the 
three interface JJs. 
These JJs are shunted by external capacitances already, unlike those in the LJJ. 
As an example of some damping we test a case where one has a lossy dielectric in the capacitor such that the effective loss tangent of the JJ 
and shunt capacitor together are $\tan\!\delta(\omega=\omega_J) = 4\cdot 10^{-3}$. 
For reference, this corresponds to parallel resistances of $\hat R_J = 1/(\tan\!\delta\, \omega_J \CJab)$
and $R_J^B =  1/(\tan\!\delta\, \omega_J \CJbb)$, 
in parallel to the shunt capacitors $\CJab$ and 
$\CJbb$ in Fig.~\ref{fig:circuit_SR}(a).    
Here we choose the LJJ frequency $\omega_J$ 
as the reference for the loss tangent 
because the interface elements temporarily oscillate 
with approximately that frequency during the gate operation.
The added damping in panel (b) give rise to slightly increased $v_f$-variations 
and slightly reduce the average $v_f$ relative to the undamped case (panel (a)). 
We believe these differences are caused because the BSR parameters 
were optimized in the absence of damping.
However the same damping after longer times, as shown in panel (c), 
diminishes $v_f$-variations, because the residual gate energy has been reduced 
significantly over time. 
In addition, this reduces the $v_f$-variations between operation types. 
This shows that the output velocities of gates can be constrained to a range 
using damping and a specified minimum delay time, regardless of the number of previous gate operations.

\begin{figure}
\includegraphics[width=0.8\columnwidth]{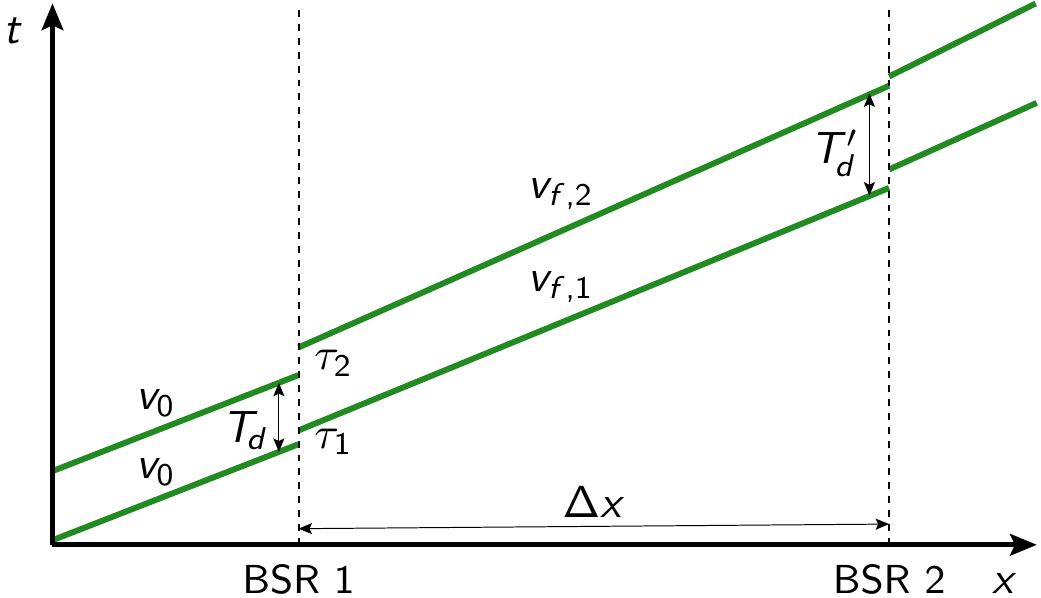}
\caption{
Sketch for operation of a 2-bit serial gate, e.g. 2-bit SISO register, 
for input of 2 successive fluxons separated by a delay time $T_d$.
The duration $\tau_j$ and output velocity $v_{f,j}$ ($j=1,2$)
of each gate operation depend on the operation type, and thus on the input state
$(\sigma_1, \sigma_2)$. 
Due to different $\tau_j$ and different $v_{f,j}$, the delay time at 
the following gate $T_d'$ will in general differ from $T_d$, 
depending on the distance $\Delta x$ between the two gates.
} 
\label{fig:sketch_Tdmod}
\end{figure}

Variability of the gate output velocities and the operation-dependent gate durations 
gives rise to timing uncertainties for the fluxon arrival at a later gate.
Take for example the 2-bit SISO register of Fig.~\ref{fig:serial2bit_SR}(a),
where we now consider two input fluxons arriving at the first BSR 
with an initial delay time $T_d$ between them, as sketched in Fig.~\ref{fig:sketch_Tdmod}.
After two bits have passed through the first BSR of the SISO, the delay time is in general modified,
due to different duration $\tau_j$ ($j=1,2$) of the two operations at that BSR. 
The fluxon delay time also changes during their subsequent motion from the first to the second BSR,
due to the different exit velocities $v_{f,j}$ ($j=1,2$) from the first BSR. 
The final fluxon delay time therefore is
$T_d' = T_d + \tau_2 - \tau_1 + \Delta x (v_{f,2}^{-1} - v_{f,1}^{-1})$, 
where $\Delta x = (N+2)a$ is the length of the LJJ section between the two BSR gates. 
Here the $\tau_j$ and $v_{f,j}$ depend on operation type and on $T_d$.
We estimate the final delay time $T_d'$ 
from the simulations of two fluxons given in Fig.~\ref{fig:variation_Tdelay}(c). 
The figure shows the output velocity of the second fluxon, $v_{f,2}$. 
The output velocity of the first fluxon is only operation-dependent, 
$v_{f,1} = 0.91 v_0$ and $0.95 v_0$ 
for NOT- and transmission-type, respectively, 
and is shown as a label in Fig.~\ref{fig:serial2bit_SR}(b-e). 
We also obtain $\tau_j$ from the simulations (not shown). 
From the different input combinations $(\sigma_1,\sigma_2)$ and samples $T_d$
we determine the mean $\langle T_d' - T_d \rangle$ and the maximum $\text{max}(T_d' - T_d)$ 
of the delay time changes as functions of $\Delta x$.
Averaging over all input combinations, the delay time remains unchanged from one BSR to the next: 
$\langle T_d' - T_d \rangle/t_{\text{LJJ}} \approx 0$, 
where $t_{\text{LJJ}} = \Delta x/v_0$ is the nominal passage time 
in the connecting LJJ and we assume $v_0=0.6 c$.
However, we find that the dominant delay-time change is caused by different input combinations. 
The maximum change in delay time is $\text{max}(T_d'-T_d)/t_{\text{LJJ}} \approx +19 \%$, 
which comes from the input combination $(\sigma_1, \sigma_2)=(-1,1)$. 
The maximum negative change in delay time is $\text{min}(T_d'-T_d)/t_{\text{LJJ}} \approx -14 \%$, 
which comes from the input combination $(\sigma_1, \sigma_2)=(1,1)$.
For the other two combinations the delay-time changes are comparably small.

As mentioned above, asynchronous gates require a minimum delay time between bits
to ensure their interactions are negligible. 
Since the delay time $T_d'$ after the gate can decrease 
as function of the LJJ length $\Delta x$
(due to one operation type dominantly) 
the LJJ should not be too long to still meet the minimum delay criterion. 
We estimate an upper limit of $\Delta x$ in the following way: 
The initial delay time $T_d$ will likely be a factor $p>1$ of the minimum delay time 
$T_{d,\text{min}} = 10 \lambda_J/v_0$.
After the gate operation, the final delay time may be decreased to 
$\text{min}(T_d') = p 10 \lambda_J/v_0 - 0.14 \Delta x/v_0$. 
Since $T_d' > T_{d,\text{min}}$ has to be fulfilled, 
one obtains the upper limit for the LJJ length of $\Delta x < 70 (p-1) \lambda_J$. 
As a lower limit for the gate-connecting LJJ length
we set $\Delta x > 10 \lambda_J$, 
i.e. the same value that also ensures negligible fluxon-fluxon interactions. 
In our experience, 
LJJs shorter than that can hinder the free fluxon motion between gates.

Another source of timing error is jitter from the LJJs, which is the fluctuation 
$\sigma_t$  
of fluxon passage times caused by thermal noise during its motion in the LJJ. 
In the ballistic regime of underdamped LJJs, 
where $\alpha \omega_J t_{\text{LJJ}} \ll 1$, 
the jitter error (standard deviation of arrival times/time) is expected to be small 
\cite{FedorovETAL2007, PanGorKuz2012}.
Herein, $\alpha$ is the damping coefficient 
of the Sine-Gordon equation \cite{ScoChuRei1976} 
and is given by the loss tangent of 
the LJJ, $\alpha = g_J/(\omega_J c_{J})$, 
with conductance $g_J$ and capacitance $c_J$ per unit length.  
In our discrete LJJs, this corresponds to the loss tangent of each JJ, 
$\alpha = \tan\!\delta =  1/(\omega_J C_J R_J)$. 
For example, in a discrete Nb-LJJ with energy scale $E_0/(k_B T) \approx 50$, 
cf.~Sec.~\ref{sec:discussion}, 
a JJ loss tangent of $\tan\!\delta = 2 \cdot 10^{-3}$ 
typical for large-area $\text{Al}\text{O}_\text{x}$ barriers, 
and fluxon speed $~0.6 c$, 
a conservative estimate for the jitter error\cite{FedorovETAL2007} 
amounts to $\sigma_t/t_{\text{LJJ}} \lesssim 2\%$ 
for motion over $(10 - 70) \lambda_J$,
corresponding to $30 - 190$ cells of our discrete LJJ.
A more refined model \cite{PanGorKuz2012} predicts even smaller jitter error. 
The jitter error for the relevant LJJ lengths is thus expected to be much 
smaller than the above-estimated timing uncertainty arising from the gate operation.
We expect thermal noise in the gate interface itself will also create jitter 
of the same order of magnitude as the LJJs. 
Therefore we did not simulate thermal noise in this work.

\section{Collective coordinate analysis}\label{sec:CCM}

For solitons and other collective excitations of a many-body system, 
the collective coordinate (CC) method is a powerful way to 
reduce the many degrees of freedom to a few essential coordinates \cite{Rajaraman}.
We have previously developed such a CC model for the fundamental 
(1-bit) RFL gates \cite{WusOsb2020_RFL}. 
Here we extend the model to the BSR, in particular the 1-input 
BSR of Fig.~\ref{fig:circuit_SR}.
To this end we parametrize the LJJ fields left and right of the
interface (at $x=0$) with the ansatz
\begin{eqnarray}
&&\phi(x<0) =
   \big(\negthinspace\phi^{(\sigma, X_L)} + \phi^{(-\sigma, -X_L)}\big)(x) 
   + 2\pi ( k_L -1 +\sigma)
\,, \nonumber \\
\label{eq:fluxoncombination_kLR}
&&\phi(x>0) =
   \big(\negthinspace\phi^{(-\sigma, X_R)} + \phi^{(\sigma, -X_R)}\big)(x) 
   + 2\pi (k_R -1)
 \,.
\end{eqnarray}
Each (left and right) field consists 
of a linear superposition of a fluxon and its mirror antifluxon, 
where $\phi^{(\sigma, X)}$ is the phase field of a fluxon 
of polarity $\sigma$
which we model as a kink equivalent to the soliton solution 
of the LJJ field \cite{Rajaraman}, 
$\phi^{(\sigma, X)}(x,t) = 4 \arctan\left( e^{-\sigma (x-X)/W} \right)$.
Herein, the time-dependent fluxon positions $X_{L,R}(t)$
serve as the dynamical coordinates of the model, 
while the fluxon width $W$ is taken to be constant 
in a so-called adiabatic approximation \cite{DauxoisPeyrard}. 
As in Sec.~\ref{sec:SR_initialization}
the integers $k_{L,R}$ describe the vacuum levels of the left and right
phase fields before the arrival of the fluxon. 
The resulting rail-JJ phase $\phiBB = \phiL-\phiR = 2\pi (k_L - k_R)$
corresponds to an initial orientation $S = (k_L-k_R)$ of the stored flux quantum.
In comparison, the CC model developed for 1-bit RFL gates \cite{WusOsb2020_RFL} 
is based on \Eq{eq:fluxoncombination_kLR} with the special case $k_L-k_R=0$.

The ansatz \eqref{eq:fluxoncombination_kLR} neglects that for $S \neq 0$
the LJJ fields may deviate in the vicinity of the interface 
from the vacuum levels, as modelled by the bound states, \Eq{eq:boundstate_kLR}. 
However, considering the relatively small bound-state amplitudes of the BSR
with $|S|=1$, this approximation seems justified.

\subsection{CC Model Parametrization and Potential}

\begin{figure*}[tb]
\includegraphics[width=\textwidth]{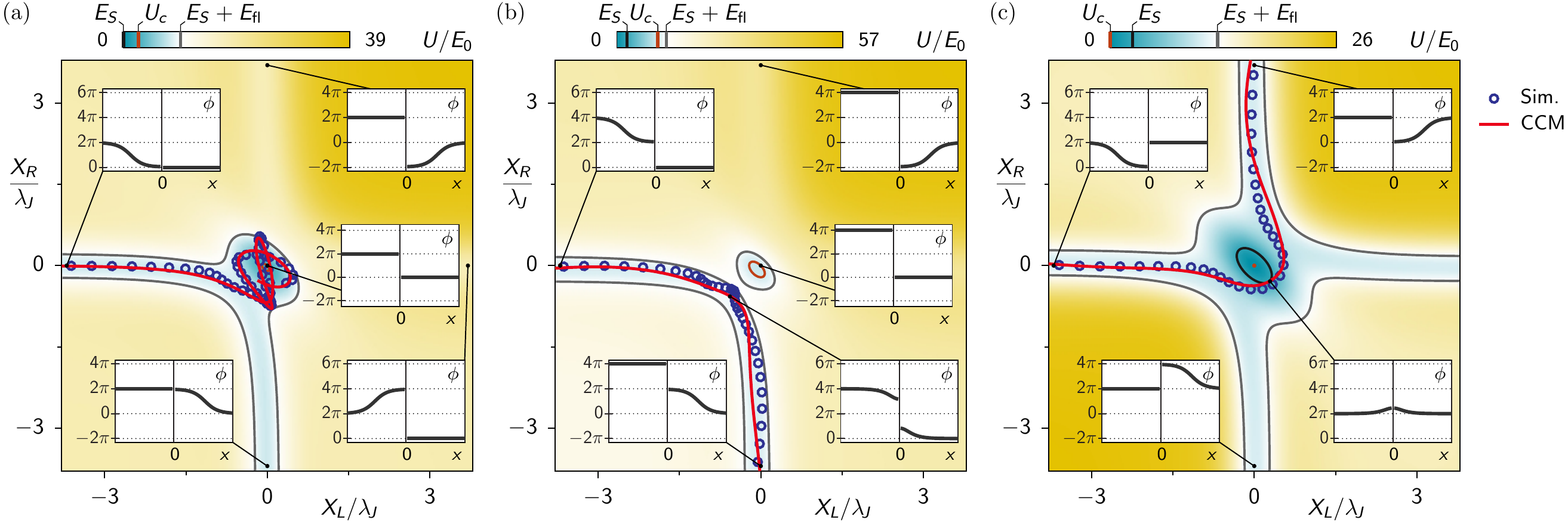}
\caption{
CC potentials $U(X_L,X_R$) and trajectories $(X_L,X_R)(t)$ (red line)
for a 1-input BSR with polarity $\sigma=1$ of incoming fluxon 
and with initally stored state (a) $S=0$, (b) $S=1$ and (c) $S=-1$. 
Equipotential lines at the following energies are shown:
stored bit energy $E_S = 2\pi^2 (L\lambda_J)/(L_s a) S^2$ (black),
initial energy $E_{\text{init}} = E_S + \Efl$ (gray),
and potential energy at center, $U(0,0)= 2\pi^2 (L\lambda_J)/(L_s a) (S + \sigma)^2$ (brown).
The CC model is based on the mirror-fluxon ansatz, \Eq{eq:fluxoncombination_kLR}, 
which is illustrated in the insets for various points $(X_L,X_R)$
in coordinate space.
The initial field distribution before fluxon arrival 
corresponds to the point $X_L \ll -\lambda_J$ and $X_R=0$ (left inset),
where the fields to the left and right of the interface are $2\pi k_L$ and $2\pi k_R$, 
with $k_L-k_R=S$. 
The CC trajectories (red) show solutions of the CC equations of motion, 
\Eq{eq:EOM_CCM0}, 
and illustrate (a) loading of the storage cell with a flux quantum (not optimized), 
(b) transmission-type BSR dynamics and (c) NOT-type BSR dynamics. 
In all cases, the CC trajectories
show good agreement with trajectories 
obtained from the full circuit simulation results $\phi^{(l,r)}_n(t)$
fit to the form of \Eq{eq:fluxoncombination_kLR} (blue markers).
The BSR parameters are those of table~\ref{tab:margins}.
Note that $U$ depends on the initial stored state $S$ and the input-fluxon polarity $\sigma$ as the product $\sigma S$,
such that for $\sigma=-1$ the potential (and dynamics) of panels (b) and (c) 
would be exchanged, while panel (a) would remain unchanged.
}
\label{fig:ccm}
\end{figure*} 

Examples for the parametrization of \Eq{eq:fluxoncombination_kLR} 
are shown in the insets of Fig.~\ref{fig:ccm}
for different points in coordinate space $(X_L,X_R)$. 
Subfigures (a-c) represent the different configurations
$(k_L,k_R)=$ $(0,0)$, $(1,0)$, and $(0,1)$, respectively,
for fluxon polarity $\sigma=1$.
A fluxon ($\sigma=1$) initially situated in the left LJJ far 
away from the interface at $X/\lambda_J \ll -1$, 
is approximated by \Eq{eq:fluxoncombination_kLR} with the coordinate 
$X_L = X$, while $X_R = 0$ describes the absence of excitations in the 
right LJJ (see left-most inset in all panels (a-c)).
For this initial state, \Eq{eq:fluxoncombination_kLR}
forms a step at the interface ($x=0$) 
between $2\pi k_L$ to the left and $2\pi k_R$ to the right, 
corresponding to the initially stored flux state $S = (k_L-k_R)$.
With $k_{L,R}$ and $\sigma$ set by the initial state,
\Eq{eq:fluxoncombination_kLR} 
fulfills the boundary conditions, 
$\phi(x\to -\infty) = 2\pi k_L +  2\pi \sigma$
and $\phi(x\to \infty) = 2\pi k_R$, for any finite $X_{L,R}$.
Under these boundary conditions, 
four different asymptotic single-fluxon states
are permitted and these
can be parametrized through suitable choice of $X_{L,R}$:
a fluxon (antifluxon) in the left LJJ is parametrized by $X_L<0$ ($X_L>0$) 
together with $X_R=0$ and
a fluxon (antifluxon) in the right LJJ is parametrized by $X_L=0$
together with $X_R<0$ ($X_R>0$). 
In the center of the coordinate space, $(X_L,X_R)=(0,0)$,
the phase distribution forms a step 
between $2\pi (k_L + \sigma)$ to the left and $2\pi k_R$ 
to the right of the interface, where
$\phiBB = 2\pi (k_L-k_R + \sigma)$. 
At points near $(X_L,X_R)\approx (0,0)$, 
the step is modified by the possible excitations near the interface, see e.g.
the inset in bottom right corner of Fig.~\ref{fig:ccm}(c).
In the corners of the configuration space where $(|X_L|\gg 1, |X_R|\gg 1)$, 
\Eq{eq:fluxoncombination_kLR} describes unavailable (high-energy) two-fluxon states
(not shown).

Using \Eq{eq:fluxoncombination_kLR} we can derive
the collective coordinate model for the BSR. 
This derivation is discussed in detail in Appendix \ref{app:CCM}, 
while here we simply summarize the results. 
After inserting \Eq{eq:fluxoncombination_kLR} 
into the system Lagrangian, \Eq{eq:Lagrangian} is simplified to 
\begin{equation}\label{eq:L_CCM}   
\frac{\mathcal{L}}{E_0} =
\frac{1}{2}
   \left(\!\!\begin{array}{c} \dot X_L \\[1ex] \dot X_R \end{array} \!\!\right)
 \mathbf{M} 
   \left(\!\!\begin{array}{c} \dot X_L \\[1ex] \dot X_R \end{array} \!\!\right)
   - U(X_L,X_R)
\end{equation}
where $U(X_L,X_R)$ is the dimensionless CC potential, 
and the mass matrix $\mathbf{M}$ is composed of the coordinate-dependent, 
dimensionless elements
$M_{ii} = m_i(X_i)$ and $M_{i,j \neq i} = m_{LR}(X_L,X_R)$ ($i,j=L,R$).
The CC potential $U$, masses $m_i$ and mass coupling $m_{LR}$ 
are given 
in \Eqs{eqA:U_CCM}, \eqref{eqA:mXi} and \eqref{eqA:mXLXR}, respectively.
Compared with the CC model of the 1-bit RFL gates \cite{WusOsb2020_RFL}, 
an additional term 
\begin{eqnarray}\label{eq:uS_CCM}
&& u_s
 = \frac{1}{2} \left( \sigma(\phiL - \phiR  + 2\pi f_E )\right)^2 
 = \frac{1}{2} \Bigl( 2\pi \sigma (k_L-k_R) \Bigr. \\
&&\quad  \bigl.+\, 8 \arctan e^{X_L/W} - 8 \arctan e^{-X_R/W} + 2\pi (1 + \sigma f_E)
 \bigr)^2 \nonumber
\end{eqnarray}
contributes to the CC potential $U$, cf.~\Eq{eqA:U_CCM},
which stems from the shunt current through the inductor $L_s$.

The diagonal elements $m_i$ of the mass matrix $\mathbf{M}$ 
vary with $X_i$ near the interface, 
but asymptotically ($|X_i| \ll \lambda_J$) 
approach a constant value, $m_i = 8\lambda_J/W$.
The mass coupling $m_{LR}$ is exponentially suppressed far away from the 
interface, but is finite near it.
It is proportional to the rail-JJ capacitance $C_J^B$,
and this explains the important role of $C_J^B$ for the forward-scattering of a fluxon 
from one LJJ to another in many gates.
In the fundamental (NOT and ID) RFL gates, mass coupling is the dominant coupling mechanism
between the LJJs, whereas coupling generated by the potential is negligible
since the potential gradient always acts 
perpendicular to the coordinate axes, $\left. \partial U/\partial X_i\right|_{X_i=0} = 0$.
Relative to the fundmental RFL gates, 
the BSR has an added contribution $u_s$ in the CC potential $U$
which can generate a much stronger coupling between $X_L$ and $X_R$, 
depending on the configuration $(k_L,k_R)$.

The CC potential $U$ is shown in Fig.~\ref{fig:ccm} for the BSR parameters
of table~\ref{tab:margins}, $\sigma=1$, 
and three different configurations $(k_L,k_R)$. 
We emphasize that $U$ 
depends parametrically on parameters of 
the initial state, namely on the initially stored SFQ, $S=k_L-k_R$, 
and on the polarity $\sigma$ of the incoming fluxon.
Specifically, the dependence enters in form of the product $\sigma \cdot S$,
as can be seen from \Eq{eq:uS_CCM} 
for the case of zero external flux through the storage cell, $f_E=0$.
All other contributions to the CC potential, 
$U_0$, $u_1$, and $u_2$ in \Eq{eqA:U_CCM},  
are independent of both $\sigma$ and $S$.
Note that the product $\sigma \cdot S$ preserves the circuit's invariance
under phase-inversion ($\sigma \to -\sigma$ and $S \to -S$), 
which would only be broken in presence of finite $f_E$.

Most contributions to the CC potential, $U_0$, $u_1$, and $u_2$ in \Eq{eqA:U_CCM},
have  mirror symmetry about the line $X_R = -X_L$,
whereas $u_s$ has this symmetry only for $\sigma(k_L-k_R) = -1$, 
as can be seen from \Eq{eq:uS_CCM} for $f_E=0$.
In Fig.~\ref{fig:ccm} the mirror symmetry is thus seen only in panel (c), 
where $\sigma (k_L-k_R)=-1$, 
while in the other cases it is broken. 
The asymmetry is particularly strong for $\sigma (k_L-k_R)=+1$, panel (b).

A fluxon initially at position $X\ll-\lambda_J$, moving with velocity $v_0$,
is parametrized by $(X_L,X_R)=(X,0)$, 
$(\dot X_L, \dot X_R) = (v_0, 0)$, and the related fluxon width
$W/\lambda_J = \sqrt{1-v_0^2/c^2}$.
For these initial conditions, 
the initial energy of the system is found from \Eq{eq:L_CCM} 
to be $E_{\text{init}} = E_S + \Efl(v_0)$.
Herein, $\Efl(v_0)$ is the initial fluxon energy, \Eq{eq:Efluxon}, and 
$E_S$ is the energy of the initially stored flux quantum $S$, 
as given in \Eq{eq:E_S__bs__ccm}.
The coordinate space accessible during free evolution 
with energy $E_{\text{init}}$ is indicated by the corresponding 
equipotential lines (gray) in Figs.~\ref{fig:ccm}(a-c).
In Fig.~\ref{fig:ccm}(c) this space consists of a central well which connects
four asymptotic `scattering valleys'. All of these correspond to a single fluxon, 
but differ by its polarity or its position in either the left or right LJJ,
cf.~description above. 
In Figs.~\ref{fig:ccm}(a,b), 
only two of these valleys are connected as a result of the potential's asymmetry, 
namely the fluxon's input valley ($X_L<0$, $X_R\approx 0$),
with the valley ($X_L\approx0$, $X_R< 0$)
that corresponds to a forward-scattered fluxon.

The Lagrangian, \Eq{eq:L_CCM}, generates the coupled equations of motion,
\begin{eqnarray}\label{eq:EOM_CCM0}
 \left(\!\!\begin{array}{c} \ddot X_L \\[1ex] \ddot X_R \end{array} \!\!\right)
= - \mathbf{M}^{-1} \left(\!\!\begin{array}{l}
 c^2 \frac{\partial U}{\partial X_L} 
 + \frac{1}{2}\frac{\partial m_L}{\partial X_L}  \dot X_L^2 
 + \frac{\partial m_{LR}}{\partial X_R} \dot X_R^2 \\[1ex]
 c^2 \frac{\partial U}{\partial X_R}
 + \frac{1}{2}\frac{\partial m_R}{\partial X_R} \dot X_R^2 
 + \frac{\partial m_{LR}}{\partial X_L} \dot X_L^2 
 \end{array} \!\!\right)
 ,\quad
\end{eqnarray}
which describe the free dynamics of the coordinates $X_{L,R}$
for fixed initial values of $S$ and $\sigma$. 
Recall that in our CC model, \Eq{eq:fluxoncombination_kLR}, 
$S$ and $\sigma$ are mere parameters determined by the initial state. 
However, as will become clearer in the discussion below, 
the corresponding values $S'$ and $\sigma'$ after the scattering
can also be deduced from the asymptotic states of the evolution.

\subsection{CC Model Results}

From \Eq{eq:EOM_CCM0} we obtain the CC trajectories $(X_L,X_R)(t)$ 
shown in Figs.~\ref{fig:ccm}(a-c) (red lines).
We also compare the CC trajectories with `simulated' trajectories
which are obtained by fitting the phases $\phi^{l,r}_n(t)$ of the 
full circuit simulations to \Eq{eq:fluxoncombination_kLR} (blue markers). 
As Fig.~\ref{fig:ccm} demonstrates, 
there is generally very good qualitative and quantitative agreement 
with the CC model trajectory. 

Next we describe how an empty BSR circuit is loaded with an SFQ such that it is
initialized for regular BSR operation. 
In Fig.~\ref{fig:ccm}(a) the trajectory of the incoming fluxon enters the 
central potential well where it bounces multiple times. 
Since no damping has been included in the CC dynamics, \Eq{eq:EOM_CCM0},
the trajectory may eventually exit the well, 
corresponding to a fluxon emitted from the interface 
into the left or right LJJ (insets). 
In the circuit simulations, however, even though no resistances
are included, the generation of plasma waves at the interface 
effectively constitutes a weak damping mechanism which prevents
later escape from the interface. 
For the coordinates trapped near the center point $(X_L,X_R)\approx (0,0)$, 
the phase distribution is close to a step profile 
(inset pointing to center of coordinate space), 
whereas the initial fluxon profile has vanished. 
From a comparison with the initial state 
(inset pointing to left side of coordinate space), 
where no flux was stored in the BSR ($S=0$), 
it is clear that a flux quantum has now been added to the storage cell, 
i.e., $S'=1= S+\sigma$.
While Fig.~\ref{fig:ccm}(a) shows the dynamics for a high-energy 
fluxon ($\dot X_L(0) = v_0 = 0.6c$) 
in absence of an external flux $f_E$ through the storage cell, 
the noise in the loading process can in principle be lowered by using a low-energy fluxon. 
This decreases the amount of energy that must be lost to capture the fluxon. 
To avoid back reflection of the low-energy fluxon from the gate interface, 
one must lower the interface potential by applying a flux $f_E \neq 0$. 

Figure \ref{fig:ccm}(b) shows results for the initial state $S=1$. 
Here the central well of $U$ is separated by a potential barrier from the 
input valley. The trajectory is shown to follow along the curved potential into 
the scattering valley, which corresponds to a fluxon in the right LJJ, 
while $S$ remains unaffected (inset).
This process thus corresponds to the transmission of a fluxon 
without topological change
-- the initial phase difference of $2\pi$ between the left and right LJJ 
is therefore roughly maintained during the scattering.
In comparison, the dynamics of an ID gate is more complicated (as well as longer
in duration), strongly relying 
on mass-coupling forces (cf.~Fig.~4(a) in Ref.~\onlinecite{WusOsb2020_RFL}).

Figure \ref{fig:ccm}(c) shows results for the BSR initially in the state $S=-1$. 
Here the CC potential 
resembles that of a fundamental 1-bit gates 
(cf.~Fig.~4(a) in Ref.~\onlinecite{WusOsb2020_RFL}).
The resulting CC dynamics is similar to the NOT fundamental gate dynamics
(cf.~Fig.~7(c1) in Ref.~\onlinecite{WusOsb2020_RFL}), and can be explained
by the combined effect of the strong mass-coupling (large $\CJbb$)
together with the forces from the potential ($\IJbb$, $L_s$) 
and the mass-gradients ($\CJbb$, $\CJab$).
The resulting state after the scattering corresponds to a forward-scattered 
antifluxon, while the stored flux has been inverted, $S' = \sigma = -S$ (inset).

As these examples show, the CC model 
-- though heavily simplifying the many-JJ circuit to a reduced system 
with two degrees of freedom -- 
describes the fluxon scattering in the BSR accurately. 
Furthermore, it is a good tool to interpret and predict fluxon dynamics at circuit interfaces.
With the help of the CC model, we were able to understand how 
the product $\sigma \cdot S$,
which represents the relative polarity of moving and stored bit states, 
changes the potential landscape. 
This in turn changes the scattering dynamics, 
as we described for the three relevant cases 
(initialization and the two distinct BSR gate operations).

\section{Discussion}\label{sec:discussion}

Shift registers \cite{SR_RSFQ, SR_AQFP, Yorozu2003, VolETAL2013, Ishida2016, BautistaETAL2022}
constitute a class of superconducting memory \cite{Brock2001, SolETAL2017} intended 
to provide fast access to small amounts of data. 
They are usually loaded as serial input and intended as first-in first-out (FIFO) buffers.
This is different than RAM, which is a larger memory intended 
for addressing in a 2D array. 
In superconducting logic, 
the currently available RAM \cite{Hidaka2006, Tolpygo2019, HerrHerr2021}
typically builds upon vortex-transition memory \cite{Tahara1995, Enomoto1999}
for SFQ-based logic. 
However, new AQFP \cite{sc_memory_QFP},  
magnetic-superconducting hybrid \cite{sc_memory_piJJ},
nanowire-based \cite{sc_memory_nanowire1, sc_memory_nanowire2},
and DRAM \cite{sc_memory_DRAM} memories 
can provide sufficient memory for near-term superconducting logic applications. 

In the future we intend to describe how to build large shift registers.
The 2-bit shift register described above is made from two 1-bit BSRs. 
It shows that ballistic gates can in principle be executed in a sequence of two
without external power with certain defined constraints. 
The same holds for the 2-input BSRs and furthermore the simulation data 
indicates that multi-bit line BSRs can be operated with similar performance.
This would allow some bit lines to be used for input and others 
to be used for sending data to the needed outputs. 
Having a logical depth of two without external power provides a useful feature 
because almost every gate is currently clocked (and not sequenced)
in most SFQ logic. 
This could also lead to new methods to incorporate memory with logic
(cf.~Ref.~\onlinecite{MajorityGateLogic2020}). 
In principle, RFL BSRs could allow greater density than shift register 
that use dual rail encoding \cite{Yorozu2003}
because RFL already uses an SFQ for both bit states. 
We finally note that, in principle,  
a dual-rail RSFQ could produce an output with both rails merged 
such that it feeds data into a BSR.

The 2-input BSRs in this work (with 2-outputs) can in principle be used 
to shift information on-demand in two dimensions (vertically and horizontally). 
In previous work, we also showed how to make a CNOT gate \cite{WusOsb2020_CNOT}, 
which can provide XOR operations. 
In the future, we plan to release gates to do full digital processing. 
The XOR operation could give the sum operation for either a half or full adder. 
To realize such an adder, we plan to study a reversible gate that executes 
the missing multiplication function.

Logic gates generally have different execution times (delays) \cite{Sutherland1991}, 
and in SFQ logic, delays and jitter lead to requirements
for clock and bit synchronization \cite{DualRailPolonsky,DualRailJapan,Sherwood2021} 
and sometimes bit arbitration  \cite{TaharaETAL2001}. 
However, some progress has been made recently in this area with the
introduction of a second time-constant within an AND and OR gate
to conveniently define a timing window for receiving the SFQ (bit=1 state) \cite{Rylov2019}. 
Additionally, our asynchronous gates provide a positive development for RFL 
(and any future bipolar-SFQ logic)
because the timing requirement basically reduces to a requirement of bit order. 
In RSFQ, the (unipolar-bit) T flip-flop is comparable in that it uses 
no clocking and the bits only need to arrive in a particular order. 

Superconducting logic can in principle approach thermodynamic reversibility, 
meaning that there is no lower limit to energy dissipation. 
In the past this was studied with so-called adiabatic logics 
\cite{parametricquantron1985, SemDanAve2003, RenSemETAL2009, RenSem2011,RQFPgate}.
In these logic types, adiabatic clock waveforms drive the gates 
such that the circuit state is always close to a potential-energy minimum. 
In these adiabatically reversible logic types the dissipated energy
scales with the inverse of the clock period 
such that the thermodynamic limit is approached by lowering the clock frequency. 
Ballistically reversible logic follows an alternative approach. 
In ballistic superconducting logic, fluxons in LJJs have been chosen as the bit carriers.
The gates are powered by the same input fluxons while no external power is applied. 
Output states with the same energy as the input states are accessible 
through the free dynamics powered by the fluxon's energy entering the gate circuit. 
Near-thermodynamic reversibility relies on successful exiting 
of the output fluxon state with a practical velocity while 
only small amounts of energy are lost to other degrees of freedom. 
BSRs, similar to previous ballistic RFL gates, satisfy these reversible criteria. 

When setting the gate energy efficiency to $\geq 86 \%$, 
we obtain the wide parameter margins shown in Table \ref{tab:margins}. 
It follows that the energy cost per operation is $E_{\text{op}}<0.14 E_B$,
where $E_B$ is the energy of the input bit (which is independent of its bit state).
For the assumed input velocity of the fluxon $v_0/c=0.6$ 
this bit energy is $E_B = 10 E_0$,
compared with the rest energy $8 E_0$ of a stationary bit (cf.~\Eq{eq:Efluxon}).
Herein, the energy scale $E_0 = I_c \Phi_0 \lambda_J/(2\pi a)$
depends on the LJJ fabrication. 
The BSR can be fabricated from digital foundry materials, 
such as Nb superconductor with an AlOx barrier, 
similar to previous RFL gates \cite{Liuqi2019}. 
For example, in an LJJ built with the discreteness used in our simulations, 
$a/\lambda_J = \sqrt{2\pi I_c L/\Phi_0} = 1/\sqrt{7}$,
and with $I_c = 3 \umA$, 
an energy cost of $< 3.7 \uzJ/\text{op}$ would result. 
When experimentally optimized and realized, 
this could compare favorably with state-of-the-art 
logic efficiency results (cf.~Ref.~\onlinecite{YoshikawaETAL2013}). 
With other materials, one could in principle lower $E_B$ closer to $k_B T$ 
to achieve an even lower energy cost.

Although it is beyond the scope of this work to specify a full architecture for RFL, 
it is obvious that the BSRs could be tested by a train of fluxons 
traveling with some interval between them. 
For example in the simulation data of Fig.~\ref{fig:phi_x_t_SR_190619__2fluxonevent}, 
the input fluxons arrive with a time interval of $T = 8/\nu_J$ for clarity 
(smaller intervals are possible). 
The energy cost estimated above from the circuit simulation includes the loss to 
plasma waves (from the imperfectly reversible undamped gates). 
At very high Josephson frequencies comparable to the superconducting gap, additional
loss due to quasiparticles is expected and would therefore 
set an upper limit to the Josephson frequency $\nu_J$ and the resulting operation speed.
Assuming circuits are made with a Josephson frequency $\nu_J=44 \uGHz$ 
and the above-mentioned time interval is used between bits, 
we calculate a real time interval of $182 \,\text{ps}/\text{op}$. 
From the rate ($5.5 \uGHz$) and the above energy calculation, 
the maximum power loss per bit (during operations) is estimated 
as $20 \,\text{pW}$. Our gates allow a combined benefit of asynchronous timing and energy efficiency. 
With these modest assumptions, the maximum energy delay product (EDP) for the 
shift register is less than $6.6 \cdot 10^{-31} \,\text{Js} = 10^3 h$. 
Clocking will cost energy as well, and will be addressed in future work. 
However, we note that an EDP on the order of $1 \cdot 10^{-22} \,\text{Js}$ has been modeled in a one-stage clocked RSFQ architecture \cite{Sherwood2021}. 
In the future, we plan to target low EDP in all of our gates, including those that are less efficient non-ballistic gates (see, e.g., Ref.~\onlinecite{WusOsb2020_CNOT} 
for a clock-triggered gate).

\section{Conclusion}

Reversible logic may progress digital computing generally because it allows 
great improvements in computing efficiency at the gate level. In contrast, 
end-of-the-roadmap CMOS will have orders of magnitude higher energy cost per bit 
switching. The type of reversible logic gates which we studied here (RFL BSRs) are ballistic. 
By introducing the asynchronous feature to ballistic gates, as we have done in this work, we expect greater 
practicality in our reversible logic family (RFL) since the timing requirements are reduced. 

The ADF (Anderson, Dynes, and Fulton) flux shuttle provided 
a pioneering design for a shift register prior to the start of SFQ logic. That logic
is thermodynamically irreversible with the bit's energy dissipated during every 
logic operation. In contrast to the ADF shuttle, which has bits encoded by SFQ presence, our RFL ballistic gates conditionally invert fluxon polarity, where the fluxon 
polarity encodes the bit state. In this work, we introduce BSRs (Ballistic Shift Registers) which add the feature of memory to previous ballistic multi-port gates. The BSRs rely only on ballistic scattering dynamics between the input fluxon and stored SFQ. Here the gate dynamics fall into two cases which consist of the resonant NOT case, generally used in RFL, and a simpler transmission case.

We have performed circuit simulations of a 1-input BSR, 
as well as a shift register composed of two 1-input BSR gates in sequence. 
In another design we introduced a 2-input BSR which can be used as a register 
with separate write and read ports or alternatively as a device to shift the bit 
state between different bit lines. Furthermore, we discuss how this may be 
helpful for a register-based memory. 

Since the ballistic scattering depends on the stored bit state (unlike previous 
RFL gates), the 1-input and 2-input BSR constitute the first set of 
asynchronous reversible logic gates appropriate for feed forward computing. The 
former (1-input) gate is shown to allow the execution of two in sequence without 
external power. The latter gate allows more bit lines to be added. We discussed 
how this is related to logical depth and timing requirements. 

Most importantly technically, perhaps, is that the BSR has wide parameter margins, 
where all margins are above $46 \%$ when the energy efficiency is set 
to 86\%. This is far above the variation in today's standard 
fabrication processes such that BSRs can be tested.

In addition to full circuit simulations, we have modeled the BSR dynamics 
by a collective coordinate model which reduces the many-JJ degrees of freedom 
to only two coordinates. With the help of this model, the 
state-dependent scattering dynamics can be understood from 
effective potentials in fluxon coordinate space -- the BSR scattering potentials 
are dependent on the initial fluxon and SFQ states. 

All SFQ logic types switch in a time equal or greater than the natural 
oscillation period of the Josephson junctions. Our logic is a fast reversible 
logic type in that it is designed to switch in only a few Josephson periods, and it has the potential to be faster than adiabatically powered reversible logic which is generally slowed by the adiabatic power source. 
Consecutive  BSR operations are shown to be possible in less than 8 Josephson 
periods. We do not currently see the need for an adiabatic clock in contrast to 
other reversible logic families and this helps enable logic at high speed. Thus, 
we are optimistic that our ballistic logic may enable high-throughput 
high-efficiency computation with unpowered gate sequences.

\section*{Acknowledgments} 
KDO would like to thank 
the Herrs, M.~Frank, R.~Lewis, N.~Missert, I.~Sutherland, V.~Semenov, K.~O'Brien, 
B.~Sarabi, C.~Richardson, D.~Mountain, G.~Herrera, and N.~Yoshikawa 
for stimulating scientific discussions.
We thank Seeqc (www.seeqc.com) for their professional foundry services 
which were used to fabricate RFL gates.
WW would like to thank the Physics Department 
at the University of Otago for its hospitality.

\begin{appendix}

\begin{figure*}
\includegraphics[width=\textwidth]{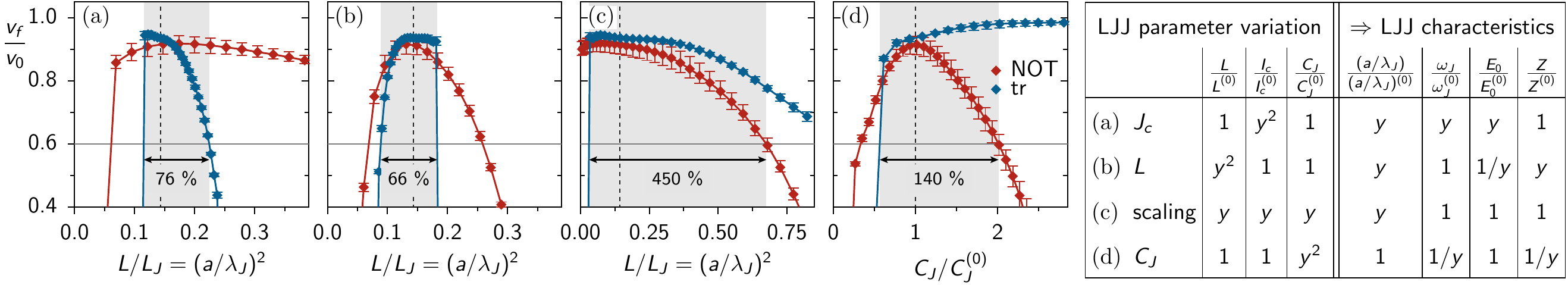}
\caption{
Margins of 1-input BSR under variation of LJJ parameters:
(a) critical current density $J_c$, (b) cell inductances $L$, (c) 
cell inductances $L$ and JJ areas, and (d) JJ capacitances $C_J$.
The graphs are constructed similar to those in Fig.~\ref{fig:margins}.
In (a-c), the x-axis shows the resulting variation of the 
relative discreteness, $(a/\lambda_J)^2 = L/L_J = 2\pi L I_c/\Phi_0$, 
instead of the actually varied parameter(s), 
while in (d) it shows the varied capacitance $C_J$ 
relative to its nominal value $C_J^{(0)}$. 
All interface parameters are kept constant at values given in Table \ref{tab:margins}.
}
\label{fig:margins_Dx} 
\end{figure*}

\section{Collective coordinate analysis}\label{app:CCM}

Here we sketch the derivation of a collective coordinate model 
for the 1-input BSR of Fig.~\ref{fig:circuit_SR}(a), leading to the results
discussed in Sec.~\ref{sec:CCM}.
The procedure is similar to the collective coordinate analysis for 
other RFL gates \cite{WusOsb2020_RFL}.

The starting point is the circuit Lagrangian, 
\Eqs{eq:Lagrangian} with the interface contribution, \Eq{eq:Lc_BB1_Lshunted}.
Inserting the mirror fluxon ansatz, \Eq{eq:fluxoncombination_kLR},
the LJJ contributions become
\begin{eqnarray}\label{eqA:L_CCM_LJJonly}
&& \frac{1}{E_0} \left(\mathcal{L}_l + \mathcal{L}_r \right)
= \sum_{i=L,R} \frac{m_0(X_i)}{2} \frac{\dot{X}_i^2}{c^2} - U_0(X_L,X_R) 
\,, \\
\label{eqA:U0_CCM_LJJonly} 
&& U_0 = \sum_{i=L,R} \Biggl\{ 
 \frac{4\lambda_J}{W} \left( 1 - \frac{2 z_i}{\sinh(2 z_i)} \right) \Biggr.\\
&&\hspace*{1.5cm} \Biggl.+ \frac{2 W}{\lambda_J} 
 \tanh(z_i) \sech^2(z_i) \left[ 2 z_i + \sinh(2 z_i) \right] \Biggr\} 
 \,, \nonumber  \\
\label{eqA:m0_CCM_LJJonly} 
&& m_0(X_i) = \frac{8\lambda_J}{W} \left(1 + \frac{2 z_i}{\sinh(2z_i)} \right)
\,,
\end{eqnarray} 
where $z_i = X_i/W$ ($i=L,R$).
To obtain these expressions, we have replaced the 
LJJ sums in $\mathcal{L}_{l,r}$ by integrals,
based on the small discreteness, $a/\lambda_J \ll 1$. 
We have evaluated all integrals with boundaries $(-\infty,0)$ and $(0,\infty)$,
which corresponds to including the interface's termination JJs as part of the LJJ.
To correct for this, the corresponding energies have to be subtracted 
in the interface Lagrangian $\mathcal{L}_I$, \Eq{eq:Lc_BB1_Lshunted},
such that $\CJab \to \CJab - C_J$ and $\IJab \to \IJab - I_c$.
After inserting the ansatz \Eq{eq:fluxoncombination_kLR} also into $\mathcal{L}_I$,
the full system Lagrangian reads
\begin{equation}\label{eqA:L_CCM}   
\frac{\mathcal{L}}{E_0} =
  \frac{m_{\!L}\dot{X}_{\!L}^2}{2c^2} + \frac{m_{\!R}\dot{X}_{\!R}^2}{2c^2} 
+ m_{\!LR} \frac{\dot{X}_{\!L} \dot{X}_{\!R}}{c^2} - U(X_{\!L},X_{\!R})  
    \,.
\end{equation}
Herein, the interface modifies 
the dimensionless mass of \Eq{eqA:m0_CCM_LJJonly} and 
also contributes a mass-coupling term, 
\begin{eqnarray}
\label{eqA:mXi}
&&m_i(X_i) = m_0(X_i) + \frac{\CJab-C_J + \CJbb}{C_J \lambda_J/a} (g_I(X_i))^2 
\,,\\
\label{eqA:mXLXR}
&&m_{\!LR}(X_L,X_R) = \frac{\CJbb}{C_J \lambda_J/a} g_I(X_L) g_I(X_R) 
\,,
\end{eqnarray}
where the factor $g_I(X_i) = 4 \left(\lambda_J/W\right) \sech(X_i/W)$
describes the local influence of the interface.
The dimensionless CC potential of \Eq{eqA:U0_CCM_LJJonly} is also modified,
\begin{equation}
\label{eqA:U_CCM}   
U = U_0 + \frac{\IJab-I_c+\IJbb}{I_c\lambda_J/a} u_1 
 + \frac{\IJbb}{I_c\lambda_J/a} u_2  
 + \frac{L \lambda_J/a}{L_s} u_s
 \,,   
\end{equation}
with the interface contributions to the potential,
\begin{align}
\label{eqA:u1_CCM}
 u_1 &= \sum_{i=L,R} 8 \sech^2(z_i) \tanh^2(z_i) 
 \,,  \\
\label{eqA:u2_CCM}
 u_2 &= - \prod_{i=L,R} \left[ 8 \sech^2(z_i) \tanh^2(z_i) \right] 
 \, \\
 &+ \prod_{i=L,R} \left[4 \sech(z_i) \tanh(z_i) \left( 1 - 2 \sech^2(z_i) \right) \right] \nonumber \\
\intertext{and} \nonumber\\ 
\label{eqA:u3_CCM}
 u_s
 &= \frac{1}{2} \left( \sigma(\phiL-\phiR) + 2\pi \sigma f_E\right)^2 
 \,.
\end{align}
Using $\phiL = \phi(x=0^{-}) = 8\arctan e^{\sigma X_L/W} + 2\pi (k_L -1 + \sigma)$
and $\phiR = \phi(x=0^{+}) = 8\arctan e^{-\sigma X_R/W} + 2\pi (k_R -1)$
from \Eq{eq:fluxoncombination_kLR},
the storage-cell contribution $u_s$ can be written as 
\begin{eqnarray}
\label{eqA:u3_CCM_v2}
 &&u_s
 = \frac{1}{2} \Bigl( 
 8 \arctan e^{X_L/W} - 8 \arctan e^{-X_R/W} \Bigr. \nonumber \\
 &&\hspace*{1.2cm} \Bigl. +\,2\pi\sigma(k_L-k_R) + 2\pi ( 1 + \sigma f_E) \Bigr)^2 
 \,.
\end{eqnarray}

\section{LJJ parameter margins}\label{app:Dx_variation}

Similar to the variation of interface parameters in Fig.~\ref{fig:margins}, 
in Fig.~\ref{fig:margins_Dx} we show the output-to-input velocity ratio under
variations of different LJJ parameters:
(a) the JJ critical currents $I_c$,
(b) the cell inductances $L$, 
(c) the cell inductances $L$ along with $I_c$ and the JJ capacitances $C_J$, 
and (d) $C_J$.
Case (a) corresponds to a variation of the critical current density $J_c$.
Case (c) can geometrically correspond to linearly scaling the inductor length 
while scaling the JJ area by the same amount 
(the dependence of the inductance on the inductor width is more complex). 
The LJJ parameter variations in general give rise to scalings of 
(i) Josephson penetration depth relative to unit cell length 
$\lambda_J/a \equiv \sqrt{L_J/L} = \sqrt{\Phi_0/2\pi L I_c}$,
(ii) JJ frequency $\omega_J=1/\sqrt{L_J C_J} = \sqrt{2\pi I_c/\Phi_0 C_J}$, 
(iii) LJJ energy scale $E_0 = (\Phi_0/2\pi)^{3/2} \sqrt{I_c/L}$, and 
(iv) LJJ impedance $Z=\sqrt{L/C_J}$, 
as shown in the table of Fig.~\ref{fig:margins_Dx}.
The first three variations (a-c) entail 
variations of the relative lattice spacing $a/\lambda_J$,
and we present the data in the respective panels 
as functions of $(a/\lambda_J)^2$. 
For most LJJ parameter variations, 
the margins of this quantity are typically around $60 - 70 \%$, 
as e.g. seen here in cases (a,b).
An exception is case (c) where all parameters of the LJJ are scaled by the same
factor, as indicated in the table. 
Here the resulting margins of $(a/\lambda_J)^2$
are almost an order of magnitude larger, and we attribute this robustness 
to the invariance of the fluxon energy $10 E_0$. 
In contrast, in cases (a) and (b) the fluxon energy is lowered relative to its
nominal value for decreasing and increasing values of $a/\lambda_J$, 
respectively. 
This leads to characteristic edges seen in the variation data of the 
transmission-type scattering (blue markers), appearing when the fluxon's initial 
kinetic energy is too low to overcome the potential barrier on the transmission path. 

We note that for large discreteness, $(a/\lambda_J)^2 \approx 1$, 
the moving fluxon loses energy through excitation of linear plasma modes 
in the LJJ \cite{WusOsb2020_RFL, BraunKivshar1998}. 
Because of the resulting strong damping, the output-to-input velocity ratio 
in panel (c) becomes increasingly ill-defined for $(a/\lambda_J)^2 \lesssim 1$
(it becomes dependent on the distance of the measurement points 
from the gate interface). 

Panel (d) shows the varied capacitance $C_J$ relative to its nominal value $C_J^{(0)}$. 
We note that all of our simulations are performed with $a/\lambda_J$ as the sole free parameter, and absolute values of $C_J^{(0)}$, $I_c^{(0)}$, and $L^{(0)}$ 
are not specified.
However, for the Nb-fabrication assumed in the discussion of Sec.~\ref{sec:discussion},
the nominal values become
$L^{(0)} = 16 \upH$, $I_c^{(0)} = 3.0 \umA$ 
and $C_J^{(0)} = 120 \ufF$.
Panel (d) shows wider $C_J$-margins ($140\%$) compared with the inductance margins 
of panels (a,b). 
We attribute this increase to the fact that the fluxon energy is not changed here, 
unlike in cases (a,b), cf.~the table in Fig.~\ref{fig:margins_Dx}.
However, the margins are also smaller than in the case of panel (c). 
In case (d) we observe that high values of $C_J$ limit the NOT operation (red markers). 
This is likely due to the lowered LJJ frequency $\omega_J$;
the NOT gate is a resonant process that depends on the frequency of the LJJs.

\end{appendix}


\end{document}